\documentstyle[prl,aps,psfig,floats,twocolumn]{revtex}   
\newcommand{\be}{\begin{equation}}      
\newcommand{\ee}{\end{equation}}      
      
\newcommand{\bef}{\begin{figure}}      
\newcommand{\eef}{\end{figure}}      
\newcommand{\bea}{\begin{eqnarray}}    
\newcommand{\eea}{\end{eqnarray}}

\newcommand{\lan}{\langle}    
\newcommand{\ran}{\rangle}

\def\spose#1{\hbox to 0pt{#1\hss}}      
\def\ltapprox{\mathrel{\spose{\lower 3pt\hbox{$\mathchar"218$}}      
 \raise 2.0pt\hbox{$\mathchar"13C$}}}      
\def\gtapprox{\mathrel{\spose{\lower 3pt\hbox{$\mathchar"218$}}      
 \raise 2.0pt\hbox{$\mathchar"13E$}}}      
\def\inapprox{\mathrel{\spose{\lower 3pt\hbox{$\mathchar"218$}}      
 \raise 2.0pt\hbox{$\mathchar"232$}}}

\newcommand{\bean}{\begin{eqnarray*}}  
\newcommand{\eean}{\end{eqnarray*}}

\def\lsim{\raise 0.4ex\hbox{$<$}\kern -0.8em\lower 0.62ex\hbox{$\sim$}}  
\def\gsim{\raise 0.4ex\hbox{$>$}\kern -0.7em\lower 0.62ex\hbox{$\sim$}}

\begin{document}  
\draft  
  
\twocolumn[\hsize\textwidth\columnwidth\hsize\csname  
@twocolumnfalse\endcsname  
  
\title{The Glass-like Universe: \\ 
Real-space correlation properties of standard  
cosmological models}  
  
%Super-homogeneous distributions and   
%the Harrison Zeldovich Power Spectrum}  

\author{Andrea Gabrielli$^{1}$,   
Michael Joyce$^{2}$ and Francesco Sylos Labini$^{2,1}$}   
\address{$^{1}$ INFM Sezione Roma1,    
Dip. di Fisica, Universit\'a ``La Sapienza'',  
P.le A. Moro, 2, I-00185 Roma, Italy.  }  
\address{$^{2}$  
LPT, Universit\'e Paris XI, B\^atiment 211, F-91405   
Orsay, France }  
  
\date{\today}  
\maketitle  
\begin{abstract}  
After reviewing the basic relevant properties  
of stationary stochastic processes (SSP), defining basic  
terms and quantities, we discuss the properties of the  
so-called Harrison-Zeldovich like spectra. These correlations, 
usually characterized exclusively in $k$-space (i.e. in terms  
of power spectra $P(k)$), are a fundamental feature of all  
current standard cosmological models. Examining them in real space 
we note their characteristics to be a {\it negative} power  
law tail $\xi(r) \sim - r^{-4}$, and a {\it sub-poissonian}  
normalised variance in spheres $\sigma^2(R) \sim R^{-4} \ln R$.   
We note in particular that this latter behaviour is at the  
limit of the most rapid decay ($\sim R^{-4}$) of this quantity  
possible for {\it any} stochastic distribution (continuous or discrete). 
This very particular characteristic is usually obscured  
in cosmology by the use of Gaussian spheres.  
In a simple classification of all SSP into three categories, we highlight  
with the name ``super-homogeneous'' the properties of the class  
to which models like this, with $P(0)=0$, belong. In statistical 
physics language they are well described as glass-like. They do not  
have either ``scale-invariant'' features, in the sense of critical  
phenomena, nor fractal properties. 
We illustrate their properties with some simple examples, in 
particular that of a ``shuffled'' lattice.  
\end{abstract}  
  
] \narrowtext  
  
\section{Introduction}  
  
In standard theories of structure formation in cosmology   
the density field in the early Universe 
is described as a perfectly homogeneous and isotropic matter  
distribution, with superimposed tiny fluctuations   
characterized by some particular correlation  
properties (e.g. \cite{pee93}). These fluctuations are believed   
to be the initial seeds from which, through a complex dynamical  
evolution,  galaxies and galaxy structures have emerged.  
In particular the initial fluctuations are taken to have 
Gaussian statistics and a spectrum which is exactly, or very  
close to, the so-called {\it Harrison-Zeldovich} (hereafter H-Z)  
\cite{har,zel} 
or ``scale-invariant'' power spectrum (hereafter PS).   
Because fluctuations are Gaussian, the knowledge of the PS, or its  
Fourier conjugate, the real space correlation function, gives 
a complete statistical description of the fluctuations. 
The H-Z type spectrum was first given a special importance 
in cosmology with arguments for its ``naturalness'' as an 
initial condition for fluctuations in the framework of  
the expanding universe cosmology, and it is in this context 
that the use of the term ``scale-invariant'' to designate it 
can be understood. It subsequently gained in importance  
with the advent of inflationary models in the eighties, and 
the demonstration that such models quite generically predict 
a spectrum of fluctuations of this type. Since the early nineties,  
when the COBE experiment \cite{cobe}   
measured for the first time fluctuations  
in the temperature in the Cosmic Microwave Background  
Radiation (hereafter CMBR) at large scales,  
and found results consistent with the predictions of models 
with a H-Z spectrum at such scales, the H-Z type spectra have  
become a central pillar of standard models of structure formation 
in the Universe. The aim of the present paper is twofold. Firstly, 
to clarify the statistical properties in real space of these  
distributions, which have been almost completely overlooked in  
the literature on the subject.  
And secondly, through this discussion, 
to relate and compare this model of the primordial Universe  
to correlated systems encountered in statistical physics.  
We attempt to make the paper as self-contained as possible, and 
not excessively technical in its discussion either of cosmology or 
statistical concepts, in the hope that it may be easily  
accessible to both cosmologists and statistical physicists.

The H-Z spectrum arises in cosmology through a particular condition 
applied to perturbations of Friedman-Robertson-Walker (FRW) models, 
which  describe a homogeneous Universe in expansion. This condition -  
commonly referred to in cosmology as ``scale invariance'' of the  
perturbations - gives rise to a spectrum (commonly called the  
``scale-invariant'' perturbation spectrum) with $P(k) \sim k$ 
at small $k$. 
All current standard cosmological models of structure formation in the 
Universe assume a spectrum exactly like this, or close to it, 
as initial condition for perturbations in the Universe.  
In such models there is at any time a finite scale corresponding  
to the causal horizon, which increases with time, and below  
which causal physics can act to modify the spectrum.  
This causal physics depends, in general, on the details of the model,  
i.e. on the nature of its content in matter and radiation (or other 
forms of energy), until a characteristic time (the time when matter  
and radiation have comparable densities), after which purely  
gravitational evolution takes over. There are many variants on  
standard cosmological models e.g. ``Cold Dark Matter'' (hereafter CDM),  
``Mixed Dark Matter'' (hereafter MDM), or the currently favored  
one with a non zero cosmological constant  
(hereafter $\Lambda$CDM), each of them leading to a different form  
for the spectrum at smaller scales (i.e. large $k$) which can be calculated.  
In CDM models (in which the predominant massive 
component driving collapse under gravity is Cold Dark Matter, ``cold''  
in the sense that the particles have little initial velocity dispersion)   
the PS decays at small scales  
(large $k$) as a negative power law in $k$, 
while in Hot Dark Matter (hereafter HDM) 
models (for which the prototype is a Universe dominated by 
a light neutrino) there is a exponential cut-off in the spectrum (due  
essentially to the fact that the ``hot'' neutrinos  
wipe out structures at these scales with their large velocity dispersion).  
All of these models, however, have the same ``primordial'' 
H-Z spectrum $P(k) \sim k$ (or very close to it) on large  
scales (i.e. small $k$), that is at scales which are large compared 
to the causal scale at the time of matter-radiation equality.  
This latter scale is of course much smaller than our present  
causal horizon (i.e. than the part of the Universe we can  
probe today). This means, in particular, that  
these primordial density correlations  
should be imprinted in the  
distribution of matter at very large scales, and should in principle  
be detectable in the distribution of galaxies at very large scales, 
inside the present horizon.  
Until now the only probe of fluctuations on such scales is through  
the temperature variations in angle of the CMBR, as 
the angular correlations in temperature fluctuations are coupled  
directly to the three dimensional density fluctuations. From the  
COBE measurements \cite{cobe}  the  amplitude of the fluctuations 
inferred is    
$\sim 10^{-5}$ in the PS at these scales. We will  
discuss elsewhere the practical difficulties involved  
in measuring such a weak signal in the discrete distribution of  
galaxies. Here we concentrate on identifying the  
real space properties of these theoretical  
models at large scales.  
  
Another context in which an understanding of the statistical  
properties in real space of the H-Z PS of the mass  
density field is important is represented by  
cosmological N-body simulations, the aim of which is to calculate the  
formation of structures under gravity in the Universe by a  
direct numerical calculation (see e.g. \cite{efst85,jenkins98}). 
Because the time scale of evolution in these simulations is short 
compared to the dynamical time of the system (i.e. a particle  
moves a small distance relative to the size of the box 
representing a large volume of the Universe) the final  
configuration depends strongly on the initial conditions  
(hereafter IC) at all but the smallest 
scales. Indeed a central idea is that from the  
final distribution - which should be closely related to 
the observed one of galaxies - one should be able to  
``reconstruct'' some important features of the IC, which can 
be related to other observations such as those of the CMBR.   
A key issue for these simulations is thus the setting up of these 
IC, which involves subtle problems concerning the discretisation 
of the system. The usual approach to this problem is again  
entirely phrased in $k$-space, where instead  
a real space approach  
proves very useful \cite{thierry}. To avoid any possible  
confusion for those somewhat familiar with these  
simulations, we note here at the outset  
that the description of the H-Z model we give in this paper, 
as lattice-like or glass-like, has no direct relation to the 
use of lattices or glasses in setting up IC in current  
N-body simulations. There lattices or glasses are understood to 
be sufficiently ``homogeneous'' configurations on which to 
superimpose fluctuations of a desired type. The reason for  
their use (instead of a ``uniform'' Poisson configuration)   
is purely numerical \cite{thierry}, 
and it has nothing to do with  
the {\it intrinsic} statistical  
properties of the systems being modeled.  
Indeed, as we will explain further at the appropriate below,  
these methods have been used primarily 
to simulate cosmological models at the smaller scales at which 
they are not at all glass-like.

Discussions of real space properties of the density fluctuations  
encountered  
in cosmology are puzzlingly  
sparse in the literature on the subject. 
Peebles briefly notes (\cite{pee93} - see pg.523)  
that a  very particular 
characteristic of H-Z models is that  
``on large scales the fluctuations have to be anti-correlated   
to suppress the root mean square mass contrast on the scale   
of the Hubble length''.  Indeed, we emphasize the fact that these models  
are characterized at large scales by a correlation function $\xi(r)$  
which has a negative power-law tail: detecting it would be 
the real space equivalent  of finding the turnover to  H-Z behavior  
to scales at which the PS  goes as $P(k) \sim k$.  The preference for  
a $k$-space description is probably rooted in the fact that the linear 
dynamics, which are used to describe many problems in cosmology, are 
most naturally treated in this space.  While it is true of course that  
this description in $k$-space is complete, this by no means implies  
that the complementary real space view is redundant, as is well known 
in many contexts in physics. One of the points of this paper is to 
show that this complementary view of these apparently so familiar  
models is at the very least interesting   
and useful, in particular in how it facilitates comparisons with 
familiar physical systems.

A basic question we try to answer is the following:  
What ``kind'' of two-point correlation function is the  
one corresponding to the H-Z behaviour in cosmological models?   
We compare it to some different statistical homogeneous 
and isotropic  systems:  
(i) Poisson-like distributions, (ii)  systems   
with a power-law correlation function found in critical phenomena  
\cite{ma84} and (iii) distributions characterized  
by long-range order (e.g. lattice or glass-like) \cite{lebowitz}.  
Through this comparison we can classify H-Z models in the third 
category.  We introduce the term ``super-homogeneous''  
to refer to these kinds of distributions, as their primary  
characteristic is that mass fluctuations decay at large scales  
faster than in a completely uncorrelated (Poisson) system.   
For critical systems one has instead a decay of  
the mass variance which is slower than Poisson.  
Formally the definition of this class of ``super-homogeneous'' distributions 
is given by the condition that the PS has $P(0)=0$, or  
equivalently in real space that the integral of the two point correlation  
function over all space is zero.  
In the cosmological literature the latter property of 
cosmological models is often noted, but its meaning (as 
a strong {\it non-local}  
condition on a stochastic process) is not appreciated, 
or worse misunderstood as a trivial condition applying to any 
correlated system. In the textbook of Padmanabhan\cite{padm}, 
for example, it is ``proved'' on pg. 171 that the integral over 
all space of the correlation function vanishes  
independently of the functional behaviour of $\xi(r)$. The error is  
in an implicit assumption made that the number of particles  
in a large volume in a single realization converges exactly 
to the ensemble average. This is not true because, in general, 
extensive quantities such as particle number have fluctuations 
which are increasing functions of the volume (e.g. Poissonian, 
for which the integral is not zero). A slightly different, 
but common, kind of misunderstanding of the meaning of the  
vanishing of the integral over the correlation function is  
evidenced in the book by Kolb \& Turner \cite{kolbturner}. 
There it is affirmed (after its statement in Eq.(9.39)) to be  
``...just a statement  of mass conservation: if galaxies are clustered on   
small scales, then on large scale they must be ``anti-clustered''   
to conserve the total amount of mass (number of galaxies)''.  
The source of this misconception seems to be a confusion with the  
so-called ``integral constraint'' in data analysis  
(e.g. \cite{pee80,saslaw}), which imposes such a condition 
on the {\it estimator} of the correlation function in a {\it finite} sample, 
due to the fact that the (unknown) average number of points in such a sample  
is estimated by the (exactly known) number of points in the actual sample.  
Despite their apparent similarity, these are different conditions:  
the first (infinite volume) integral constraint provides non-trivial 
physical information about the intrinsic probabilistic nature   
of fluctuations, while the second is just an artifact of the  
boundary conditions which holds in a finite sample independently 
of the nature of the underlying correlations.  
We will discuss this point in a 
little more detail at the appropriate point below.

The paper is organized as it follows. In the first section we recall  
the basic properties of mass distributions  
(both continuous and discrete) described in terms of  
stationary stochastic processes with a well defined  
(non-zero) average density. 
In this context we introduce the basic statistical quantities   
(homogeneity scale, correlation functions, real space mass variance,  
PS etc.) used to describe these systems. We discuss  
in particular the relation between the mass variance in spheres 
and the PS, noting that for power law spectra 
$P(k) \sim k^n$ and $n \geq 1$, the small scale (i.e. large $k$) 
power dominates the real space variance at any scale.  
We explain that this is 
not a simple mathematical pathology but corresponds to a real 
property of these distributions. Indeed for discrete  
distributions of points we note that  a theorem 
has been proved \cite{kendall} showing the behaviour approached at 
$n=1$ to be the limiting decay of the variance, in real space 
spheres, in any distribution. 
In the subsequent section we discuss the H-Z spectrum, recalling 
the construction which leads to it in cosmology and why it is 
called ``scale-invariant''. We note that the H-Z criterion, as 
naively understood, is not one which is satisfied exactly by 
the spectrum $P(k) \sim k$. In the next section we give a  
classification of all SSP in terms of the behaviour of the 
PS as $k \rightarrow 0$. We give the name  
``super-homogeneous'' to those which have $P(0)=0$, 
referring to their basic characteristic as more homogeneous 
than the Poisson distribution, with a sub-poissonian  
decay of their mass variance.   In the following section 
we give the examples of a lattice, and then a  
``shuffled'' lattice,  to illustrate the properties  
of distributions of this type, which have the strong 
order of a lattice or glass at large scales. Here we 
discuss also briefly the relation of our description to 
N-body simulations. In the final section we discuss  
various points in summarizing our findings. In 
particular we clarify the use of the term  
``scale-invariance'', ``fractal'' and  
``correlation length'' 
in relation to the H-Z  
spectrum in cosmology.

\section{Basic Statistical Properties and Concepts}  
\label{basics} 
 
Inhomogeneities in cosmology are described using the general 
framework of stationary stochastic processes (hereafter SSP).  
Let us consider in general the description of a continuous or a discrete mass  
distribution $\rho(\vec{r})$ in terms of such a process. A stochastic  
process is completely characterized by its ``probability density functional''  
${\cal P}[\rho(\vec{r})]$ which gives the probability that  
the result of the stochastic process is the density field  
$\rho(\vec{r})$ (e.g. see Gaussian  
functional distributions \cite{Dobs}). For a discrete  
mass distribution  
the space (e.g. infinite three dimensional space) is divided 
into sufficiently small cells and the stochastic process  
consists in occupying or not any cell with a  
point-particle, and $\rho(\vec{r})$ can be  
written in general as:  
\be   
 \label{shi1a}   
 \rho(\vec{r})=\sum_{i=1}^{\infty} \delta(\vec{r} -\vec{r}_i)\,,          
 \ee      
where $\vec{r}_i$ is the position vector of the particle $i$   
of the distribution .

The stationarity refers in the present context to spatial stationarity  
of the process, and means that the functional   
${\cal P}[\rho(\vec{r})]$ is invariant under spatial translation.  
This property is also called the {\em statistical homogeneity} of the   
distribution.   
We suppose also that the distribution is {\em statistically isotropic}   
(invariance of ${\cal P}[\rho(\vec{r})]$ under spatial rotation),  
and has a well defined average value $\rho_0$, that is  
\be  
\label{shi2a}  
\lan \rho(\vec{r}) \ran = \rho_0 > 0\,,   
\ee  
where $\lan ... \ran$ is the ensemble average over all the possible   
realizations of the stochastic process, i.e. the average over the functional   
${\cal P}[\rho(\vec{r})]$.  
Statistical homogeneity and isotropy (hereafter  
SHI) 
imply that the $l$-point correlation functions   
$\left<\rho(\vec{r}_1)...\rho(\vec{r}_l)\right>$, for any $l$, depend  
only on the scalar relative distances among the $l$ points \cite{gsl01}.     
Moreover we assume that ${\cal P}[\rho(\vec{r})]$ is {\em ergodic}.  
In order to clarify the meaning of {\em ergodicity}, let us take  
a generic observable $F=F(\rho(\vec{r}_1),\rho(\vec{r}_2),...)$ of the  
local density 
$\rho(\vec{r})$.  
Ergodicity means that $\left<F\right>$ is equal to the spatial average  
$\overline{F}$ given by:  
\be  
\overline{F}=  
\frac{1}{\|\Omega\|}\int_{\Omega}   
d^3r_0 F(\rho(\vec{r}_1-\vec{r}_0),\rho(\vec{r}_2-\vec{r}_0),...)  
\label{ergo}  
\ee  
where the integral  
is extended to the whole space $\Omega$ and $\|\Omega\|$ is   
its (infinite) volume,   
and where $\rho(\vec{r})$ is (almost) any realization  
of the particle distribution  
``extracted'' from the functional ${\cal P}[\rho(\vec{r})]$.   
This property is also referred as the {\em self-averaging} property  
of the distribution.   
Note that if the average in Eq.~(\ref{ergo}) is extended only   
to a finite sub-sample  
$V$ of the whole space $\Omega$ , then Eq.~(\ref{ergo}) is only    
an {\em estimator} of $\left<F\right>$ in the given sub-sample.

In a single realization of the mass distribution    
the existence of a well defined average density implies that      
\cite{gsl01}  
\be     
\label{shi1}    
\lim_{R\rightarrow\infty}     
\frac{1}{\|C(R;\vec{x}_0)\|}\int_{C(R,\vec{x}_0)}d^3r\,   
\rho(\vec{r})=\rho_0>0    
\ee      
where $\|C(R,\vec{x}_0)\| \equiv 4\pi R^3/3$ is the volume of a sphere      
$C(R,\vec{x}_0)$ of radius $R$,  centered on the  
{\it arbitrary} point $\vec{x}_0$ of space  
\footnote{Because of the arbitrariness of the  
position of the center of the sphere,  
the average density is a  
one-point statistical property.}.    
When Eq.\ref{shi1} is valid one can then define \cite{gsl01} a  
characteristic {\em homogeneity scale} as  
the scale $\lambda_0$ given by  
\be    
\label{shi2}    
\left| \frac{1}{C(R;\vec{x}_0)}   
\int_{C(R;\vec{x}_0)}d^3r\, \rho(\vec{r})-\rho_0 \right| < \rho_0    
\;\;\forall R>\lambda_0,\,\forall \vec{x}_0  
 \ee      
which depends on the nature of the fluctuations of the density in 
spheres. In practice, in characterizing the scale at which a system  
begins to be homogeneous,  
it is easier to use directly some simple two-point  
statistics. We will mention these definitions at the appropriate point below.

The quantity $\left<\rho(\vec{r_1})\rho(\vec{r_2})...\rho(\vec{r_l})\right>$  
is called the        
{\em complete} $l$-point correlation function. In the discrete  
case $\left<\rho(\vec{r_1})\rho(\vec{r_2})...\rho(\vec{r_l})    
\right>dV_1,dV_2,...,dV_l$ gives the {\em a priori} probability of finding   
$l$ particles, in a single realization, placed        
in the infinitesimal volumes $dV_1,dV_2,...,dV_l$ respectively       
around $\vec{r_1}, \vec{r_2},...,\vec{r_l}$.

Let us analyze in further detail the auto-correlation properties of these   
systems. Due to the hypothesis of statistical homogeneity and isotropy,        
$\left<\rho(\vec{r_1})\rho(\vec{r_2})\right>$ depends only on       
$r_{12}=|\vec{r_1}-\vec{r_2}|$.       
Moreover,       
$\left<\rho(\vec{r_1})
\rho(\vec{r_2})\rho(\vec{r_3})\right>$ is only a function of        
$r_{12}=| \vec{r_1}-\vec{r_2}|$, $r_{23}=| \vec{r_2}-\vec{r_3}|$ and       
$r_{13}=|\vec{r_1}-\vec{r_3}|$.       
The {\em reduced} two and three-point correlation functions 
$\tilde\xi(r)$      
and $\tilde\zeta(r_{12},r_{23},r_{13})$ are respectively defined by:     
\bea     
\label{shi4}    
&&\left<\rho(\vec{r_1})    
\rho(\vec{r_2})\right> \equiv   
\rho_0^2\left[1+\tilde\xi(r_{12})\right]      
\\    
&&\left<\rho(\vec{r_1})    
\rho(\vec{r_2})\rho(\vec{r_3})\right> \equiv    
\rho_0^3      
\left[1+\tilde\xi(r_{12})+\tilde\xi(r_{23})+\right.\nonumber\\      
&&\left.\tilde\xi(r_{13})+\tilde\zeta(r_{12},r_{23},r_{13}) \right]\,.  
\label{shi4b}  
\eea     
The correlation function $\tilde \xi(r)$ is one way
to measure the "persitence of memory" of spatial variations
in the mass density \cite{huang}. 
Note that, as shown more explicitly  
below, in the discrete case the functions $\tilde \xi$ and $\tilde \zeta$  
differ from the usual $\xi$ and $\zeta$ used in cosmology  
by the so-called {\em diagonal part}.

In the discrete case of particle distributions it is very important   
to consider observations from a point occupied by a particle.  
In order to characterize statistically these observations  
it is necessary to define another kind of average:       
the {\em conditional} average $\left<F \right>_p$.       
This is defined as an ensemble average with the       
condition that the origin of the coordinates is an occupied point 
\cite{gsl01}.      
When only one realization $\rho(\vec{r})$ extracted from   
${\cal P}[\rho(\vec{r})]$  
is available,  $\left<F \right>_p$     
can be substituted by the spatial average:  
\be  
\overline{F(\rho(\vec{r}_1),\rho(\vec{r}_2),...)}_p=\frac{1}{N}\!  
\sum_{i=1}^{N}\!  
F(\rho(\vec{r}_1 + \vec{r'}_i), \rho(\vec{r}_2 + \vec{r'}_i),...)\,  
\label{ergo2}  
\ee  
where the sum is restricted to all the points ($N\rightarrow\infty$)  
$\vec{r'}_i$ occupied by a particle of the distribution.   
Again in the case in which the average is restricted to the particles  
$\vec{r'}_i$ belonging to a finite sample of volume $V$ of the whole space,  
we can consider Eq.(\ref{ergo2}) only as an {\em estimator} of    
$\left<F \right>_p$.

The quantity    
\be    
\label{shi5}    
\left<\rho(\vec{r_1})\rho(\vec{r_2})...    
\rho(\vec{r_l})\right>_pdV_1 dV_2...dV_l       
\ee    
gives    
the average probability of finding $l$ particles placed        
in the infinitesimal volumes $dV_1,dV_2,...,dV_l$       
respectively around $\vec{r_1}, \vec{r_2},       
...,\vec{r_l}$ with the {\it condition} that       
the origin of coordinates is an {\it occupied}     
point.       
We call $\left<\rho(\vec{r_1})\rho(\vec{r_2})
...\rho(\vec{r_l})\right>_p$       
conditional $l$-point density.       
    
Applying the rules of conditional probability   
one has \cite{gsl01}:            
\bea     
\label{shi6}     
&&\langle \rho(\vec{r}) \rangle_p      
=\frac{\langle \rho(\vec{0})\rho(\vec{r})\rangle}{\rho_0}\\     
&&\langle \rho(\vec{r_1})\rho(\vec{r_2}) \rangle_p =      
\frac{\langle \rho(\vec{0})  
\rho(\vec{r_1})\rho(\vec{r_2})\rangle}{\rho_0}\,.     
\nonumber     
\eea     
      
However, in general, the following convention is assumed     
in the definition of the conditional densities:       
the particle at the origin does not observe itself.       
Therefore $\langle \rho(\vec{r}) \rangle_p$ is defined only for        
$r>0$, and $\langle \rho(\vec{r_1})\rho(\vec{r_2}) \rangle_p$      
for $r_1, r_2>0$.        
Consequently, and this is what is usually done in  
cosmology \cite{pee80}, one can redefine the   
reduced two and three-point correlation function   
$\xi(r)$ and $\zeta(r_1,r_2,r_{12})$  
to be equal to $\tilde\xi$ and $\tilde\zeta$ respectively for $r,r_1,r_2>0$,  
and equal to zero for $r,r_1,r_2=0$. This means simply that the diagonal  
part is removed from $\tilde\xi$ and $\tilde\zeta$.  
In the following we use this convention.

%%%%%%%%%%%%%%%%%%%%%%%%%%%%%%%%%%%%%%%%%%%%%%%%%%%%%%%%%%%%%%%%%%  

Let us consider the paradigm of a stochastic homogeneous   
point-mass distribution: the {\it Poisson case}.  
For such a particle  distribution the reduced two-point correlation     
function Eq. (\ref{shi4})  can be written as (see \cite{gsl01})    
\be    
\label{p6}    
\tilde\xi(r) =   
\frac{\delta(\vec{r})}{\rho_0} \; \;\;(\mbox{i.e.}\;\,\xi(r)=0)\,.  
\ee    
Analogously, one can obtain the three point correlation function  
(Eq.\ref{shi4b}):       
\be    
\label{p7}    
\tilde\zeta(r_1,r_2,r_{12})=     
\frac{\delta(\vec{r_1})\delta(\vec{r_2})}{\rho_0^2}\;\;\;\;(\mbox{i.e.} \,\;  
\zeta(r_1,r_2,r_{12})=0)\,  .  
\ee       
The two previous relations are direct consequences of the fact      
that there is no correlation        
between different spatial points. That is, the reduced       
correlation functions $\tilde\xi$       
and $\tilde\zeta$ have only the diagonal part.       
The latter is present in the reduced correlation functions of any        
statistically homogeneous discrete distribution of particles with correlations.  
  
As already mentioned, in the definition of conditional densities,        
we exclude the contribution of the origin of coordinates.       
Consequently, for a Poisson distribution,  
we obtain from Eq.~(\ref{shi6}):        
\bea     
\label{p8}     
&&\langle \rho(\vec{r})\rangle_p=\rho_0 \\      
&&\langle \rho(\vec{r_1})\rho(\vec{r_2})\rangle_p=\rho_0^2      
\left[1+\frac{\delta(\vec{r_1}-\vec{r_2})}{\rho_0}\right]\;.     
\nonumber     
\eea

%%%%%%%%%%%%%%%%%%%%%%%%%%%%%%%%%%%%%%%%%%%%%%%%%%%%%%%%%%%%%%%%%%%%  

In general \cite{landau,saslaw} for a  
%statistically isotropic and homogeneous  
SHI distribution of particles the reduced    
correlation function can be written as    
\bea    
\label{pc1}   
&&\tilde\xi(r)=\frac{\delta(\vec{r})}{\rho_0}+\xi(r)\,,\\  
&&\tilde\zeta (r_1,r_2,r_{12})=     
\frac{\delta(\vec{r_1})\delta(\vec{r_2})}{\rho_0^2} +  
\zeta (r_1,r_2,r_{12})\nonumber  
 \eea       
where $\xi$ and $\zeta$ are the non-diagonal   
parts which are meaningful only for $r>0$ ad $r_1,r_2>0$ respectively.      
In general $\xi(r)$ is a smooth function of $r$  
\cite{landau,gsl01}.  
Hence we obtain from Eq.\ref{shi6}    
(by excluding  again   
the contribution of the origin of coordinates):        
 \bea     
 \label{pc2}     
 &&\langle \rho(\vec{r})\rangle_p=\rho_0[1+\xi(r)] \\      
 &&\langle \rho(\vec{r_1})\rho(\vec{r_2})\rangle_p=\rho_0^2      
 \left[1+\xi(r_{1})+\xi(r_{2})+\nonumber\right.\\      
 &&\left.\tilde \xi(r_{12})+\zeta(r_{1},r_{2},r_{12})    
 \right]\,.     
 \nonumber     
 \eea     
  
%%%%%%%%%%%%%%%%%%%%%%%%%%%%%%%%%%%%%%%%%%%%%%%%%%%%%%%%  
  
\subsection{The mass variance in a sphere}

In this section we consider the amplitude of the mass fluctuations   
in a generic sphere of radius $R$ with respect to the average mass.   
First let  $M(R)=\int_{C(R)}\rho(\vec{r}) d^3r$  
be the mass (for a discrete distribution  
the number of particles) inside the sphere $C(R)$ of radius $R$  
(and then volume $\|C(R)\|=\frac{4\pi}{3}R^3$).   
The normalised mass variance is defined as     
\begin{equation}     
\sigma^2(R)=      
\frac{\langle M(R)^2 \rangle - \langle M(R) \rangle^2}   
{\langle M(R) \rangle^2}\,,     
\label{v1}     
\end{equation}     
where     
\begin{equation}     
\langle M(R) \rangle =    
 \int_{C(R)}d^3r \langle \rho(\vec{r}) \rangle=\rho_0\|C(R)\| \,,     
\label{v2}     
\end{equation}       
and    
\begin{equation}     
\langle M(R)^2 \rangle =    
\int_{C(R)} d^3r_1\int_{C(R)}d^3r_2 \langle  
\rho(\vec{r}_1)\rho(\vec{r}_2) \rangle\;.  
\label{v3}     
\end{equation}   
Note that there is no condition on the location of the center of the   
sphere, because of the assumed translational invariance of   
${\cal P}[\rho(\vec{r})]$.  
   
In the discrete Poisson case, using Eq. (\ref{p6}), we obtain that    
\be    
\label{v4}    
\sigma^2(R) = \frac{1}{\rho_0\|C(R)\|}\equiv \frac{1}{\left<M(R)\right>}  \;.  
\ee    
  
In general, for a   
%statistically homogeneous    and isotropic  
SHI mass density field with     
correlations, substituting Eq.(\ref{shi4})   
in Eq.(\ref{v1}), we obtain  
\be    
\label{v6}    
\sigma^2(R) =  
\frac{1}{\|C(R)\|^2} \int_{C(R)} d^3r_1\int_{C(R)} d^3r_2   
\tilde\xi(|\vec{r_1} - \vec{r_2}|)  \;.  
\ee    
Using Eq.~(\ref{pc1}) in the discrete case we can write  
\bea    
&&\sigma^2(R) = \frac{1}{\rho_0\|C(R)\|}+     
\nonumber\\  
&&\frac{1}{\|C(R)\|^2} \int_{C(R)} d^3r_1\int_{C(R)} d^3r_2   
\xi(|\vec{r_1} - \vec{r_2}|)  \;.  
\label{v6b} 
\eea    
Note that the sign of the second term of Eq.(\ref{v6b}) is not uniquely   
determined. We clarify this point later on.  
Equations (\ref{v6}) and (\ref{v6b}) make evident the relation between   
fluctuations in one-point properties (as in this case the number of   
points in a sphere centered on a random point in space)  
and two-point correlations.  
In general similar links can be found between fluctuations in $n$-point  
properties and $n+1$-point correlations.  
  
Equation (\ref{shi1}) is equivalent to the requirement that  
\be  
\label{v6a}  
\lim_{R \rightarrow \infty}  
\sigma^2(R) = 0 \;, 
\ee  
which is therefore a condition satisfied by any SHI  
distribution.  
An alternative (slightly different) definition to that given  
by (\ref{shi2}) for the scale characterizing homogeneity is 
thus the scale at which $\sigma^2(R)$ reaches unity (or some other 
appropriate fiducial value)  
\footnote{Note that such a definition 
holds for SHI distributions, and not at all for the case 
of fractal systems \cite{slmp98} as discussed in our  
conclusions section below}.  
In the cosmological literature  
on the distribution of matter (galaxies, clusters  etc.) in  
the Universe there is no global 
convention about how this scale is defined; in fact it is 
a scale which is almost never discussed in precise terms. 
The two most commonly used quantities used in characterizing  
the two-point properties are (i) the scale 
\footnote{This scale has unfortunately been commonly referred to 
in the cosmological literature as the ``correlation length'' 
\cite{pee80}. It has 
no relation to the statistical physics use of the same term, which 
is a scale characterizing the rate of decay of fluctuations, not their 
amplitude. See \cite{gaite,gsld00} for a clear discussion of this point.} 
$r_0$ defined by $\xi(r_0)=1$, and (ii) the amplitude of the mass  
variance at a fiducial physical scale, taken to be $8h^{-1}$Mpc 
(e.g \cite{dp83}).  
Given (or having determined) the dependence on scale of the  
correlation function or mass variance, these can be easily related 
to simple definitions of the homogeneity  
scale. A practical working definition 
of homogeneity scale applicable in the analysis of galaxy surveys, and 
a discussion of the current status of this scale is given  
in \cite{joycesylos_ApJ2000,slmp98}.

Let us return to further discussion of Eqs.(\ref{v6}) and (\ref{v6b}).   
It is very important for our discussion to note that  
this condition (\ref{v6a}) which holds for any mass  
distribution generated by a SSP, is very different  
from the requirement   
\be  
\label{superhomo}  
\int_{\Omega} d^3r   \tilde \xi(r) = 0  
\ee  
(where $\Omega$ is the whole space)  
which is a much stronger special condition which holds for   
certain distributions - those to which below we will ascribe 
the name  ``super-homogeneous''.

Note that,  
in cosmology (e.g.\cite{pee80,saslaw}) the following approximation is often 
used 
\be    
\label{v8}    
\int_{C(R)}\!\!d^3r_1\int_{C(R)}\!\!d^3r_2   
\tilde\xi(|\vec{r_1} - \vec{r_2}|)    
\approx \|C(R)\| \int_{C(R)}\!\! d^3r   
\tilde\xi(\vec{r}) \;.  
\ee  
in particular in evaluating the variance through Eq.(\ref{v6}).     
%Using this approximation one would infer that Eq.(\ref{v6}) and  
%(\ref{v6a}) 
%actually imply (\ref{superhomo}).  
Such an approximation is not always valid,
and the convergence properties of the double integral need
to be examined carefully to establish it. In particular
it does not hold when the condition Eq. (\ref{superhomo}) 
is satisfied. This will be evident following the analysis 
we give below, as we will discuss that one has, for any
distribution (continuous or discrete), a large distance behavior 
%faster than poissonian (or ``sub-poissonian''), while
% We  
%will also discuss the result that, for  
%{\it any} stationary stochastic mass distribution  
% (discrete and continuous),  
$\sigma^2(R)=R^{-a}$ where $a\le d+1$ 
(where $d$ is the space dimension). Using the
approximation (\ref{v8}) one could apparently obtain 
through Eq.(\ref{v6}) arbitrarily rapidly decaying 
behaviors with an appropriate power-law behaviour 
in the correlation function.
  
In the discrete case, to measure $\sigma^2(R)$ one has   
to take into account both terms in Eq.(\ref{v6b}),  
not only the second one.  
From Eq.(\ref{v6b}) the variance can, in general,  
be written as the sum of two contributions:    
\be    
\label{v9}    
\sigma^2(R) = \sigma^2_{Poi}(R) + \Xi(R)  \;,  
\ee    
where the first term $\sigma^2_{Poi}$    
represents the intrinsic Poisson noise of any stochastic     
particle distribution 
\footnote{Note that this term can give a contribution 
to the variance which dominates over that due to the 
intrinsic correlations.},   
and the second term $\Xi(r)$ (which, as noted above, does not  
have to be of a determined sign) is the additional contribution due to   
correlations (i.e. to $\xi(r)\ne 0$).

%%%%%%%%%%%%%%%%%%%%%%%%%%%%%%%%%%%%%%%%%%%%%%%%%%%%%%%%%%%%%%%%%%%  

\subsection{The power spectrum}

The PS $P(\vec{k})$ is the main statistical tool used to describe  
cosmological models. It is defined as 
\be 
P(\vec{k})=\left<|\delta_\rho(\vec{k})|^2\right> 
\label{ps-defn} 
\ee 
where $\delta_\rho(\vec{k})$ is the Fourier Transform  
(hereafter FT) of the normalized  
fluctuation field $(\rho(\vec{r})-\rho_0)/\rho_0$.  
For a spatially stationary mass distribution $\rho(\vec{r})$ it is  
possible to demonstrate that it can be obtained by simply taking   
the FT of the correlation function  
$\tilde\xi(\vec{r})$ (up to a multiplicative constant) \cite{feller}:  
\be  
\label{lat7a}  
P(\vec{k}) = \frac{1}{(2\pi)^d} \int_{\Omega} 
d^dr \exp(- i \vec{k}\vec{r} ) \tilde  
\xi(\vec{r})\,.  
\ee  
Further, given statistical isotropy $P(\vec{k})\equiv P(k)$.  
For a continuous mass density field obtained by a SSP,  
the two basic properties of the PS  are  
the  following (Khintchine theorem \cite{Dobs}):  
 
1) $P(\vec{k})\ge 0\;\;\forall \vec{k}$;  
 
2) $P(\vec{k})$ is integrable in the whole space.  
  
For a discrete particle distribution the first property is still   
valid, while the second is not because of the diagonal   
part of $\tilde\xi(\vec{r})$ (the Dirac delta function in  
$\vec{r}=\vec{0}$). Indeed, this part gives a positive constant  
contribution for every $\vec{k}$ which makes the integral of  
$P(\vec{k})$ divergent. This constant contribution is   
the PS for the uncorrelated Poisson distribution of particles.  
Consequently, for discrete distributions, the property 2) is 
modified as follows: 
 
2') The FT of the $\xi(\vec{r})$ (i.e. $\tilde\xi(\vec{r})$  
without the diagonal part) is integrable in the whole $k$-space.  
 
In $d$-dimensions the properties 2) and 2') imply that:  
\bea  
\label{ps-cond1}  
&&\lim_{k\rightarrow 0} k^d P(\vec{k})=0\\  
&&\lim_{k\rightarrow\infty}k^dP(\vec{k})=0 \;.  
\label{ps-cond2}  
\eea  
where in the discrete case $P(\vec{k})$ is the  
FT of $\xi(r)$ rather than of $\tilde \xi(r)$.  
 
In three dimensions we have therefore that, in general, the PS  
can diverge as $k \rightarrow 0$ with only the condition  
that the divergence is slower that $k^{-3}$. Any standard type  
cosmological model has in this limit the H-Z spectrum, or 
something close to it, and in any case always has $P(0)=0$, 
which implies that Eq.(\ref{superhomo})  
holds. We will discuss the meaning of this condition at length below.

%%%%%%%%%%%%%%%%%%%%%%%%%%%%%%%%%%%%%%%%%%%%%%%%%%%%%%%%%%%%%%%%%%%%%%%%% 

\subsection{The PS and real space variance}  
\label{PS-variance} 
 
Let us analyze the relation between the PS and the mass-variance 
in real space. We first discuss continuous density fields, and 
then make some relevant comments on the discrete case.  
We first rewrite Eqs.~(\ref{v1}-\ref{v3}), generalizing them to the case in   
which we calculate the mass variance in a topologically more complex volume  
${\cal V}$ of size $V$. To do this one introduces the window function  
$W_{\cal V}(\vec{r})$ defined as  
\be  
W_{\cal V}(\vec{r})=\left\{  
\begin{array}{ll}  
1\;\;\;\;\mbox{if}\;\vec{r}\in {\cal V}\\  
0\;\;\;\;\;\mbox{otherwise} \;.  
\end{array}  
\right.  
\label{window}  
\ee  
Therefore we can rewrite Eq.~(\ref{v2}) as  
\begin{equation}     
\left<M({\cal V})\right>= \int_{\Omega}  W_{\cal V}(\vec{r})   
\left<\rho(\vec{r})\right> d^3 r\,.     
\label{mass2}     
\end{equation}  
and Eq.~(\ref{v3}) as  
\be  
\left<M^2({\cal V})\right>=\int_{\Omega}\int_{\Omega} 
d^3r_1 d^3r_2 W_{\cal V}(\vec{r}_1)  
W_{\cal V}(\vec{r}_2)  
\left<\rho(\vec{r}_1)\rho(\vec{r}_2)\right>\,,  
\label{v3bis}  
\ee  
where the integrals are over all space.  
The normalised variance is then given by 
\be  
\sigma^2({\cal V})=\frac{1}{V^2}\int_{\Omega}
\int_{\Omega} d^3r_1 d^3r_2   
W_{\cal V}(\vec{r}_1)W_{\cal V}(\vec{r}_2)\tilde\xi(\vec{r}_1-\vec{r}_2)\,.  
\label{sigma-w}  
\ee  
On taking the FT  one obtains  
\begin{equation}     
\sigma ^2({\cal V})=   
\frac{1}{(2\pi)^3} \int d^3 k P(\vec{k}) |\tilde{W}_{\cal V}(\vec{k})|^2     
\label{variance-ps}     
\end{equation}    
which is  explicitly positive, and $\tilde{W}_{\cal V}(\vec{k})$ is     
the FT of $W_{\cal V}(\vec{r})$, normalised by  the  
volume defined by the     
window function itself,         
\begin{equation}     
\tilde{W}_{\cal V}(\vec{k})=\frac{1}{V}     
\int_{\Omega} d^3r e^{-i \vec{k}.\vec{r}} W_{\cal V}(\vec{r})     
\label{wfn-ft}     
\end{equation}     
with $V=\int_{\Omega} W_{\cal V}(\vec{r}) d^3 r$.

Consider now again the real sphere of radius  
$R$ for which the FT of the window function     
(normalised as defined) is     
\begin{equation}     
\tilde{W}_R(\vec{k})=\frac{3}{(kR)^3} 
\left( \sin kR - kR \cos kR \right) \;.    
\label{wfn-sphere}     
\end{equation}     
One then has, assuming statistical isotropy so that $P(\vec{k})=P(k)$, 
an expression for the variance in real spheres which is  
\begin{equation}     
\sigma ^2(R)=\frac{1}{2\pi^2} \int_0^\infty\! dk      
\frac{9}{(kR)^6}\left( \sin kR - kR \cos kR \right)^2 k^2 P(k)    \;. 
\label{sigma}     
\end{equation}     

We now show that, for power-law spectra $P(k) \sim k^n$ 
(for small $k$, $n > -3$) the integral in (\ref{sigma}) has a  
very different behaviour for $n<1$ and $n \geq 1$.  
For $n<1$ the integral is dominated by $k \sim R^{-1}$, 
while for $n \geq 1$ it becomes dominated by the 
large $k$ behaviour, and therefore sensitive to the  
PS at a scale $k$ unrelated (in general)  
to $R^{-1}$. Correspondingly $\sigma ^2(R)$ is found to  
have a limiting rapidity of decay at $1/R^4$, related 
to the appearance of this divergence. Then we give the 
physical interpretation of this result, and note the 
danger of the use of a ``Gaussian window'' to mask it 
in cosmology.

So let us return to Eq.(\ref{sigma}) and take a  
PS $P(k)= A k^ne^{-k/k_c}$ (where $A$ and $k_c$ are two
constants). We consider 
$n>-3$ and take the cut-off to satisfy the convergence 
properties of the Khintchine theorem. It is easy to 
check subsequently that the results we derive are 
not sensitive to the form of this cut-off at large $k$. 
It is convenient to rescale variables to rewrite 
Eq.(\ref{sigma}) as 
\begin{equation}     
\sigma ^2(R)= \frac{9A}{2\pi^2} \frac{1}{R^{3+n}}\int_0^\infty dx      
\left( \sin x - x \cos x \right)^2  x^{n-4} e^{-x/x_{c}}     
\label{sigma-rescaled}     
\end{equation}     
with $x_c=k_cR$.  
 
Since the window function goes to unity when  $x \sim 0$ the integral 
will be well behaved at its lower limit (since $n > -3$). It is 
also convergent at its upper limit because of the exponential. 
Let us consider the dependence on the latter for a sphere 
much larger than the cut-off scale, i.e.  
$R \gg k_c^{-1}$, that is  $x_c\gg 1$.  
For $x \gg 1$ the integrand goes as $x^{n-2} \cos^2 x$,  
so that the integral converges, without the exponential  
cut-off, for $n< 1$. Thus the variance as a function 
of radius behaves as $1/R^{3+n}$, and the integral is dominated  
by modes $k \sim R^{-1}$. In fact, since the integral is independent 
of the cut-off, we have, up to a numerical factor of order unity,  
the relation  
\be     
\sigma^2(R) \approx \frac{1}{2} P(k) k^3|_{k =R^{-1}}     
\label{ps-variance} 
\ee   
so that the amplitude of the PS at $k$ can be thought 
simply to correspond to the variance at the physical scale $R^{-1}$. 
 
For $n \geq 1$, on the other-hand, the integral diverges and the cut-off 
comes into play.  For $n=1$ the integral is      
\begin{equation}     
\int \frac{dx}{x}e^{-x/x_{c}} \sim \ln x_c \sim \ln R      
\end{equation}     
so that $\sigma^2(R) \sim (\ln R)/R^{4}$.  
Finally for $n>1$ the  integral goes as $\sim x_c^{n-1}$ so  
that one gets $\sigma^2(R) \sim 1/R^{4}$, independently of $n$.  
Importantly, for $n \geq 1$, the integral in Eq.~(\ref{sigma-rescaled})     
is dominated by the short wavelengths with $k \sim k_c^{-1}$, and not     
by the fluctuations on the scale $ k \sim R^{-1}$, and correspondingly 
the relation (\ref{ps-variance}) does not hold.  
The amplitude of the PS is no longer related to the real space 
fluctuations at the scale $k \sim R^{-1}$;  
{\it instead large scale spatial 
fluctuations have their behaviour determined by the short scale power 
in the theory. }   
   
To summarize clearly: For a power-law  
$P(k) \sim k^n$ (with an appropriate cut-off around 
the wavenumber  $k_c$) the mass variance for real spheres with radius  
$R \gg k_c$ is given by  
 
\begin{enumerate} 
\item For $n<1$,  $\sigma^2(R) \sim 1/R^{3+n}$ and the dominant 
contribution comes from the PS modes at $k \sim R^{-1}$. 
\item For $n>1$, $\sigma^2(R) \sim 1/R^{4}$ and the dominant 
contribution comes from the PS modes at $k_c^{-1}$.   
\item For $n=1$, we have the limiting logarithmic divergence 
with $\sigma^2(R) \sim (\ln R)/R^{4}$. 
\end{enumerate}

In the cosmological literature 
\footnote{See, for example, the section entitled 
``Problems with filters'' in the book by Lucchin and Coles  
\cite{coles-lucchin}.} 
the divergences in the latter two  
cases are treated as a simple mathematical pathology due to the 
assumption of a perfect sphere (with a perfectly defined boundary). 
Replacing the real sphere with a smooth Gaussian filter  
$W_{\cal V}(\vec{r}) \sim e^{-r^2/R^2}$ these integrals are also 
cut-off at the scale $k \sim R^{-1}$ and one recovers a behaviour 
$\sigma^2 (R) \sim 1/R^{3+n}$ and a relation of the form  
(\ref{ps-variance}). While of course this is valid mathematically 
it misses an important point, which is that this limiting behaviour  
of the variance (as $1/R^4$) has a very real physical meaning  
which has to do with the nature of systems with such a rapidly 
decaying PS. They correspond to extremely homogeneous 
systems (i.e. extremely ordered systems) in which the variance 
really is dominated by the small scale fluctuations. Let us 
explain this point further.    
 
Firstly, that the behaviour has nothing to do in principle with 
the ideality of the perfect sphere is easily seen by considering 
a more realistic modeling of the sphere, using a window function 
giving a smearing on a length scale corresponding to the uncertainty  
in the radius of the sphere (This could correspond,  
for example, to the uncertainty in the distance measure to a galaxy). 
If it is larger than the intrinsic cut-off scale in the  
power spectrum, it is this scale which then provides the cut-off  
in the integral giving the mass variance in the sphere. Since 
this scale is in principle independent of the radius of  
the sphere $R$, the same limiting $1/R^4$ behaviour of the 
variance is recovered. 
%found for this realistic modelling of the sphere 
%(without the perfectly sharp edge).  
Thus it is a physical result for a continuous SSP  
that {\it the mass variance measured in spheres  
of radius $R$ cannot decrease faster than $1/R^4$.}

A more intuitive understanding of this fact can be gained by considering 
discrete distributions. One would reason that any continuous distribution 
can be arbitrarily well approximated at large scales by an appropriate  
discretisation process, and that therefore the same result may hold of  
discrete distributions. 
In fact such a result has  
been proved several years ago \cite{beck}: 
In $d$-dimensions there exists no discrete distribution of points  
in which the variance in spheres decays faster than $1/R^{d+1}$.  
One can see roughly why this is so by considering the most ordered 
distribution of points one might think of: a simple cubic lattice. 
The variance in a sphere is given by averaging over spheres with 
center anywhere in the unit cell. As the sphere moves in the unit 
cell the variance, one would guess (correctly!),  
in the number of points is proportional to the difference in the volume 
of the spheres, which is proportional to the surface area of spheres,  
i.e. $\propto R^{d-1}$ in $d$-dimensions. Thus the normalised variance 
scales as $1/R^{d+1}$, a result proved in rigorously in  
\cite{kendall} (see also \cite{beck} for a more  
general discussion of the problem).   
As we will discuss further below the  
regular lattice, or rather a randomized version of it, can be thought 
of as a kind of prototype for the class of distributions to which 
the H-Z spectrum belongs. They are distributions which are highly 
ordered (``glass-like'') in which the fluctuations in real space  
actually are at small scales (those at which the PS is 
cut-off). Because of this it is one of their characteristics,  
as we have seen, that {\it there is no direct relation between the  
PS  at scale $k$ and the physical variance in real space  
at the scale $R \sim k^{-1}$}.  
 
The Gaussian sphere completely obscures this behaviour  
for $n \geq 1$, giving an apparent  
behaviour of a real space variance $\propto 1/R^{d+n}$.   
It does this because it models the edge of the sphere as 
smeared on the length scale of the radius (i.e. assumes that 
the uncertainty in our measure of distance to a point is  
necessarily of order the distance). Instead of the 
dependence of the mass variance in spheres (with some  
intrinsic uncertainty in the definition of their edges)  
on the radius, the  
Gaussian sphere gives us the behaviour of the variance  
in spheres as a function of radius when both the radius  
and the smearing imposed on the edge change together  
and in linear proportion. This is a real  
space measure which one can define to recover $k$-space 
properties, but it loses completely the essential 
characteristic of the real space variance (measured 
in real physical spheres).  
 
\section{The H-Z spectrum and its real space properties}  
\label{H-Zsection} 
  
Let us first recall the kind of argument\footnote{We choose  
here a particular (but commonly used) way of describing the  
HZ spectrum which allows us to avoid too much extra formalism.  
For a commonly used formulation preferred by many cosmologists, 
in terms of a constant ``gauge independent'' potential, see  
for example \cite{LiddleLyth}.}  
that singles out 
the H-Z spectrum in cosmology, and why the term ``scale-invariant'' 
is applied to it. In a homogeneous Friedman-Robertson-Walker 
(hereafter FRW)  cosmology there is  
a fundamental characteristic length scale, the horizon scale 
$R_H(t)$. It is simply the distance light can travel from 
the Big Bang singularity $t=0$ until any given time in the 
evolution of the Universe, and it grows linearly with time. 
The H-Z criterion can be written 
\be 
\sigma_M^2 (R=R_H(t)) = {\rm constant}, 
\label{H-Z-criterion} 
\ee 
i.e it requires that {\it the mass variance at  
the horizon scale be constant}. Equivalently, 
given the proportionality of gravitational 
potential to mass, it can be stated as the constraint  
that the variance in the gravitational potential be  
constant at the horizon scale. It arises naturally 
in the framework of FRW cosmology as a kind of 
consistency constraint: the FRW is a cosmological 
solution for a homogeneous Universe, about which 
fluctuations represent an inhomogeneous perturbation. 
If we take any other prescription other than  
Eq. (\ref{H-Z-criterion}) such a description will 
always break down in the past or future, as the 
amplitude of the perturbations become arbitrarily 
large or small. It is in this specific sense that 
the resulting PS is said to be ``scale-invariant'': 
there is no characteristic scale 
at which fluctuations become large (or small), 
or put another way, they have the same amplitude  
as a function of the only scale in the model. 
As we will discuss further  
%at the appropriate point 
below, it has nothing to do with the same term as  
understood in statistical physics. There  
scale invariance is a characterization not of  
the amplitude of fluctuations, but rather is associated  
to a particular range of power-law behaviors in the  
correlation function.

More precisely the form of the H-Z spectrum is arrived at  
from the condition (\ref{H-Z-criterion}) in the following 
way. We move necessarily to a $k$ space description, 
as we need to include the dynamical evolution of the  
density field to infer the PS inside the 
horizon today. Let $\delta_{k}(t)$ be  
the amplitude of the Fourier component of the density  
contrast as a function of time. To every such mode  
$k$ we can associate a time $t_c$ at which it ``enters 
the horizon'', i.e. at which the wavelength $k^{-1}$ 
is equal in size to the horizon. Here we work  
(as almost always in cosmology) with a $k$ which is  
the FT with respect to the spatial coordinates  
which do not change with the expansion, the  
so-called ``comoving'' coordinates.  In these  
coordinates the time at which the mode enters 
the horizon is given by $k \eta =1$ where  
$\eta$ is the so-called ``conformal'' time 
given by $\eta=\int dt/a(t)$, with $a(t)$  
the scale factor describing the expansion of all 
physical scales in the Universe. (The horizon 
scale is simply $R_H(t) = a(t) \eta$, corresponding 
to horizon crossing criterion $(k/a)R_H(t)=1$.)  
The PS today (at $t=t_o$, say) 
given by $|\delta_{k}(t_o)|^2$ can be written 
in term of the amplitude of each mode $k$ when 
it entered the horizon. In linear perturbation 
theory, in the matter dominated Universe  
(i.e. recent epochs), the mode evolves as   
\be 
\delta_{k}(t_o) = \left(\frac{a(t_o)}{a(t)}\right) \delta_{k}(t) \;. 
\label{dc-scaling} 
\ee 
In the matter dominated FRW cosmology we have 
$a\propto t^{2/3}$ and thus $\eta \propto t^{1/3}$, so 
that the time $t_c(k)$ when the mode $k$ crosses  
the horizon follows $t_c(k) \propto 1/k^3$ and 
therefore  
\be 
\delta_{k}(t_o) \propto k^2 \delta_{k}(t_c) \;. 
\label{dc-scalingb} 
\ee 
The H-Z choice for the primordial PS  
$|\delta_{k}(t_o)|^2 \propto k$ is then  
singled out by imposing the criterion 
\be 
k^3 |\delta_{k}(t_c)|^2 = {\rm constant} \;, 
\label{dc-scalingc} 
\ee 
which is identified as the mass variance at the horizon scale 
$\eta = k^{-1}$. We note immediately, following the preceding  
discussion, that the latter identification is in fact valid  
only for power spectra $k^n$ with $n< 1$. Strictly speaking  
therefore it is impossible to satisfy the H-Z criterion as it 
is understood naively; or, to put it another way, the H-Z  
spectrum, that which satisfies (\ref{dc-scalingc}), does not  
satisfy the condition of ``scale invariance'' since the mass 
variance at the horizon scale ($\propto \eta$) is dominated  
in this case by the power at the cut-off scale, not by the  
modes $k \sim \eta^{-1}$.  
Taking a spectrum $k^{1-\epsilon}$ ($\epsilon>0$) one can 
get arbitrarily close to satisfying the H-Z criterion, but  
the condition of ``scale invariance'' (in the sense  
just explained) is not physically satisfiable.  
To avoid this conclusion the criterion could be refined to be 
that the mass variance in Gaussian spheres of radius of 
the horizon size be constant. While it does allow a 
mathematically coherent formulation, from a physical point 
of view it is an artificial way of avoiding the problem, 
which is that the variance at a given real space scale 
has nothing to do in principle with the amplitude of 
the PS at the inverse scale for $n \ge 1$.  
This is, as we have 
discussed in the previous section, a real physical property  
of such systems, not a mathematical artifact.  
%In the  
%conclusions we will return to this point and to 
%the danger of the use of Gaussian spheres.  
 
The H-Z spectrum can equivalently be characterised in
term of fluctuations in the gravitational potential,  
$\delta\phi(\vec{r})$, which are linked to the  
density fluctuations $\delta\rho(\vec{r})$  
via the gravitational Poisson equation 
\footnote{We simplify here to Newtonian gravity, which  
becomes a good approximation on sub-horizon scales. The 
comments given below can however be generalized to a 
rigourous formulation of perturbations to a FRW model.}:  
\begin{equation}  
\nabla^2\delta\phi(\vec{r})=-4\pi G \delta\rho(\vec{r})\,.  
\label{poii}  
\end{equation}  
From this, transformed to Fourier space, it follows that  
the PS of the potential  
$P_{\phi}(k)=\left<|\delta\hat\phi(\vec{k})|^2\right>$   
is related to the density  
PS $P(k)$ as:  
\[P_{\phi}(k)\sim \frac{P(k)}{k^4}\,.\]  
The H-Z spectrum corresponds therefore to 
$P_{\phi}(k) \propto k^{-3}$; or, considering the variance 
in real space spheres of the gravitational potential fluctuations, 
which as for the density fluctuations 
is related to the PS by Eq.(\ref{ps-variance}), one finds 
that this variance is constant as a function of $R$.
This is the alternative form in which the 
H-Z condition is often formulated. Note that the
Khintchine theorem  (cf. Eq.~(\ref{ps-cond1})) 
requires that a well defined SSP have $P(k)\sim k^a$ 
with $a> -3$ for $k\rightarrow 0$,
so that the H-Z corresponds to the limiting (disallowed)
behaviour. Equivalently the constancy of the variance 
is in contradiction with Eq.(\ref{v6a}) which requires that 
the asymptotic variance be zero (in order to have a well 
defined mean about which fluctuations are defined). The
H-Z spectrum can thus be seen as the (disallowed) limiting 
behaviour for the potential fluctuations to be treatable as an SSP.
That such a treatment be applicable to the potential fluctuations
is however not a physical requirement. The work of 
Chandrasekhar \cite{chandra43} (and see also \cite{gslp99})  
treats the gravitational force probability distribution in a point  
distribution and, in particular, shows it is well defined even in 
the Poissonian case, for which the potential fluctuations are not
an SSP ($n=0$). To treat the force field as an SSP requires only
the weaker condition $P(k) \sim k^a$ with $a>-1$.  
     
\subsection{The real space correlation function of CDM/HDM models}  
 
All of the current ``viable'' standard type cosmological models  
have a ``primordial'' PS which is the H-Z one (or 
very close to it) down to some arbitrarily small scale.  
During cosmological evolution causal physics modifies this  
spectrum at large $k$, which is roughly the causal horizon at 
that time. Around the time at which the matter in the  
Universe (with density scaling as $1/a^3$) begins to  
dominate over the radiation (with density scaling as $1/a^4$),  
the evolution becomes purely gravitational at all but the 
very smallest scales, while prior to this time it depends 
strongly on the details of the particular model. As a result 
all such models are H-Z for $k< k_{eq}$, but ``turn-over'' at this 
scale to a PS  
decreasing as a function of $k$. The form of the spectrum  
in this region depends on the details of the particular  
model. Since the scale $k_{eq}^{-1}$, being the size  
of the causal horizon at this time of matter-radiation equality, 
is much smaller than the causal horizon today, the primordial 
H-Z PS  is in principle detectable today. Indirect evidence for 
its reality come from the measurements of temperature  
fluctuations in the CMBR, which show a dependence 
on angular scale quite consistent with the H-Z spectrum, with 
the power-law spectrum  $P(k) \sim k^n$ giving a fit in 
the range $n= 1.1 \pm 0.5$ \cite{cobe}. The search to 
observe this ``turn-over'' to H-Z behaviour directly in three 
dimensions in the distribution of matter at large scales 
- a central prediction and check on such models - has so 
far proved elusive, because of weak statistics at large scales 
in observations of the distribution of galaxies. It is anticipated 
that forthcoming surveys, now being made \cite{SDSS} or close to 
completion \cite{2dF}, will have the capacity to detect this 
turn-over (see the discussion at the conclusions).  
In the cosmological literature this question is  
again treated almost exclusively in $k$ space. Here we look at the 
characteristic real space features which should be found in 
these galaxy surveys if the underlying behaviour is H-Z. In 
further works we will discuss in detail the question of the 
detection of these features, here we concentrate solely on  
their identification.  
 
We consider first the two point correlation function. In general 
the FT of the PS of standard cosmological models must be done 
numerically. Before doing so for some standard models we considered 
a simple PS which can be transformed, a H-Z spectrum with a simple 
exponential cut-off:   
\be     
\label{exem1}     
P(k) = A\times k\times e^{-\frac{k}{k_c}}    \;, 
\ee     
where $A$ is the amplitude and $k_c^{-1}$ the cut-off scale.
The correlation function is given by (see e.g. \cite{pee93})  
\be     
\label{exem2}     
\tilde\xi(r) = \frac{A}{\pi^2} \frac{\left(\frac{3}{k_c^2}-r^2\right)}     
{\left(\frac{1}{k_c^2}+r^2\right)^3}    \;. 
\ee     
For $r < r_c \equiv k_c^{-1}$ we have  
$\tilde\xi(r) \simeq \frac{3A}{\pi}k_c^4>0$, changing  
at $r \sim r_c$ to an asymptotic behaviour $\xi(r) \sim -r^{-4}$. 
Note that the correlation does not oscillate, its only zero 
crossing being at scale $r= \sqrt{3} r_c$. Simply because of 
the condition $P(0)=0$, which implies that the integral of  
the correlation function must be zero, the  correlation 
function must change sign and in this case it only does 
so once and thus remains negative at large scales.

In the normalised mass variance $\sigma^2(R)$ shows  
a corresponding change in behaviour from 
being approximately constant at small scales  
$R<r_c$ to a $\ln R / R^{4}$ decay at large scales,  
as was shown in Section(\ref{PS-variance}) above. 
Note that, unlike for the  
variance in spheres discussed in Section(\ref{PS-variance}), 
there is no limit to the rapidity of the decay of  
the correlation function (for the more general expression  
see \cite{pee80}). 
Despite the weakness of this correlation at large scales, 
however, the variance in spheres does not behave like  
that of a Poisson system, because of the  
balance between positive correlations  at small and negative  
at large scales imposed by the non-local condition $P(0)=0$.

In cosmological HDM  models the form of 
the PS is almost the same as we have just considered 
with an exponential cut-off \cite{padm}   
\be  
P(k) \sim k \exp(-k/k_c)^{3/2}  \;.
\ee  
A numerical integration verifies that the correlation function 
is essentially unchanged. 
 
\begin{figure}[tbp]  
\centerline{  
\psfig{file=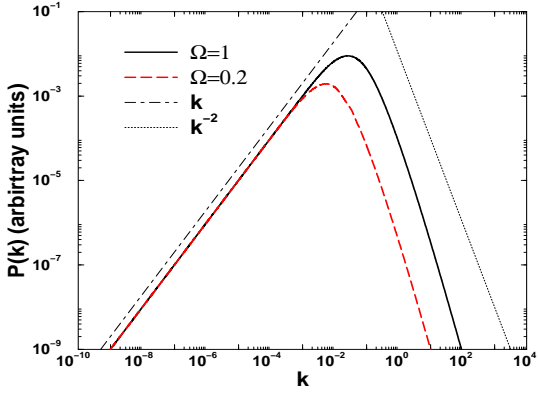,height=6cm,width=8cm,angle=0}} 
\caption{Behaviour of the power spectrum for  
a CDM model with $\Omega=1,0.2$ respectively.  
The two reference lines have exponents $k \; ,k^{-2}$.} 
\label{ps}  
\end{figure}

\begin{figure}[tbp]  
\centerline{  
\psfig{file=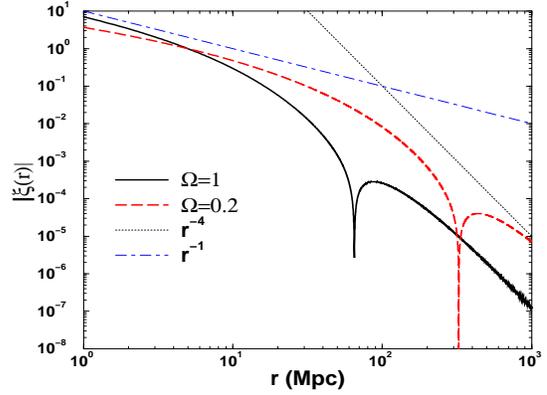,height=6cm,width=8cm,angle=0}}  
\caption{Behaviour of absolute value   
of the real space correlation  
function for the two CDM models $\Omega=1,0.2$ and $h=0.5$.   
%For convergence reasons we have chosen $k_{min}=1e-13$ and $k_{max}=1e2$.  
The two reference lines are $r^{-4}$ and $r^{-1}$. Note that at small scale  
$\tilde \xi(r) >0$, with a zero crossing at a scale depending on the location 
of the peak or ``turn-over'' in the PS, after which it remains negative  
($\tilde \xi(r) \sim -r^{-4}$) at larger distances.  
The correlation function has been normalised to be $\tilde \xi(r_0)=1$  
for $r_0=5 Mpc$.}   
\label{xi}  
\end{figure}

\begin{figure}[tbp]  
\centerline{  
\psfig{file=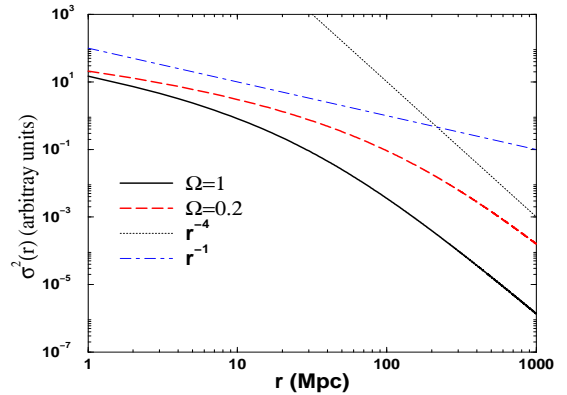,height=6cm,width=8cm,angle=0}    
}  
\caption{Behaviour of   
the unconditional variance in spheres   
for the two CDM models $\Omega=1,0.2$ and $h=0.5$.  
The two vertical lines show the transition to the   
$\xi(r) \sim r^{-4}$ behavior for the two models.   
The $r^{-4}$ behavior is a clear and distinctive feature   
corresponding to the $P(k) \sim k$ behavior.}   
\label{sigma2}  
\end{figure}

For CDM  models, the class by far favored 
in the last few years, the form of the PS  at scales  
below turn-over from H-Z behaviour is considerably more  
complicated. In a linear analysis the PS  
of CDM matter density field decays below the turn-over with a 
power-law $\sim k^{-3/2}$ at large $k$  
until a smaller scale at which  
it is cut-off with an exponential (in a manner similar  
to that in the HDM model). Numerical studies of these  
models designed to include the non-linear evolution  
bring further modifications, roughly increasing the 
exponent in the negative power law regime. For our 
analysis we have taken an analytical approximation  
to the final PS given by  
Eisenstien \& Hu \cite{ew98}, and computed  
numerically the FT to the two-point correlation function.  
We have also computed directly the variance in spheres. 
This form of the PS is given in terms of 
the various cosmological parameters. Here we consider 
for simplicity the case with the small baryon density  
set to zero ($\Omega_b=0$), which gives a PS without 
the famous oscillations reportedly detected in recent 
observations of the CMBR \cite{boomerang,maxima}. 
This structure is not of primary interest to us here 
because it can modify the correlation function only 
at small scales (it arises from causal physics  
at early times). 
In Figs.\ref{ps}-\ref{xi} we show respectively   
the behaviour of the PS and of the correlation function   
for two quite different values of the total matter  
density of the model $\Omega=1,0.2$.  
Minor differences will result in the case that 
there is a cosmological  
constant $\Omega_{\Lambda} \ne 0$ \cite{ew98}. 
In Fig.\ref{sigma2} we show the behaviour of the unconditional  
variance, computed in real-space spheres. We see again a clear 
convergence in both models to the predicted $1/R^4$  
behaviour beyond the scale characterizing the ``turn-over''.  
 
In conclusion two simple real space characteristics today in the 
distribution of matter coming from the primordial H-Z PS 
are a {\it negative non-oscillating power-law tail in the two point  
correlation function $\xi(r) \sim -r^{-4}$, and a $(\ln R)/R^4$ decay  
in the variance of mass in spheres of radius $R$}.   
These are the peculiar distinctive feature of H-Z type 
spectra which should possibly be detected in real space by  
the new galaxy catalogs.

\section{Real space classification of long range fluctuations}  
\label{classifications} 
  
We now return to a the discussion of the nature of correlations 
in systems with H-Z like power spectra, with the aim of elucidating 
their properties by comparison with systems described in  
statistical physics.  To this end we first introduce here 
a classification of all possible mass distributions in terms 
of the main features of the correlation function $\tilde \xi(r)$.  
Following from the discussion of Section(\ref{basics}) 
concerning the behaviour of mass fluctuations,  
we define three distinct classes  
(for either the case of discrete particle distribution and  
of a continuous density field):    
\begin{enumerate}  
  
\item   
\label{1.} If    
\be  
\int_{\Omega}  d^d r \; \; \tilde \xi(r) = const.>0   
\label{poi}  
\ee   
we can say that at large scale the system is   
{\em substantially Poissonian}. Indeed Eq.~(\ref{poi})  
implies that the PS goes to a constant non-zero 
value as $k$ goes to zero, and therefore that the 
large distance behavior of the mass fluctuations is  
\be  
\langle M^2(R) \rangle -\langle M(R)\rangle^2   
\sim R^d\sim \langle M(R) \rangle\,.  
\label{poi2}  
\ee  
We write here the unnormalized form of the variance, as the result 
that the variance of an extensive quantity such as the mass is  
proportional to the volume on which it is measured is the most 
intuitive way of characterizing a Poisson type behaviour.   
In this class is, for example, a system with a finite range  
correlation $\xi(r) \sim e^{-r/r_c}$. Beyond the scale $r_c$  
(the correlation length - see below 
for a discussion about this length)  
the system is uncorrelated and effectively 
Poissonian.  
 
\item \label{2.} If   
\be  
\int_{\Omega} d^d r \; \; \tilde \xi(r) = + \infty   
\label{cri}  
\ee  
then we are in a case similar to a system at the critical point  
of a second order phase transition (e.g. the liquid-gas critical point).  
Such systems have a positive correlation function which is   
asymptotically a positive power law, with $\xi(r) \sim 1/r^{\gamma}$ and 
$\gamma < d$, corresponding to a PS $P(k) \sim k^{\gamma-d}$ 
as $k \rightarrow 0$. One then has at large scales the variance 
\be  
\langle M^2(R) \rangle -\langle M(R)\rangle^2   
\sim R^{\alpha}\;\; \mbox{with}\;d \le \alpha < 2d\;,  
\label{cri2}  
\ee  
or   $\langle M^2(R) \rangle -\langle M(R)\rangle^2\sim  
\langle M(R)\rangle^{\beta}$ with $\beta=\alpha/d>1$.  
This means that mass fluctuations are large (always overwhelming 
the Poisson fluctuations) and thus they are strongly correlated 
at all scales\footnote{For example these properties near the critical point of  
the liquid-gas transition gives place to opalescence phenomena.}.  
It is in this context that the concept of self-similarity and 
scale-invariance has been introduced in statistical mechanics.  
These terms refer to the fact that in these systems the mass  
fluctuation field has well defined fractal properties \cite{slmp98}.

\item \label{3.} If  
\be  
\int_{\Omega} d^d r \; \;\tilde  \xi(r) = 0  
\label{super}  
\ee  
then, as we have discussed, we have for the behaviour of the  
mass fluctuations 
\be  
\langle M^2(R) \rangle -\langle M(R)\rangle^2   
\sim R^{\alpha}\;\;\mbox{with}\;  
d-1<\alpha<d\,,    
\label{super2}  
\ee   
i.e. $ \langle M^2(R) \rangle -\langle M(R)\rangle^2  \sim   
\langle M(R)\rangle^{\beta}$ with $\beta=\alpha/d<1$, 
so that the mass fluctuations are always asymptotically 
smaller than in the uncorrelated Poisson case. This 
also corresponds to a strongly correlated, long-range ordered, 
system. We will refer to them with the term ``super-homogeneous'' 
to underline this feature that they are more homogeneous than a  
Poisson system. (Indeed, the Poisson particle distribution is  
considered as the paradigm  of a stochastic homogeneous mass  
distribution \cite{feller}). In the context of statistical mechanics  
they can be described as glass-like, as they have the properties 
of glasses, which are highly ordered compact systems. That can be said 
to be typically lattice-like, with a long-range ordered packing, but 
without the discrete symmetries of an exact lattice. Note again that,  
since $\tilde\xi(0)>0$ (a Dirac delta function in the discrete case)  
by definition, $\tilde\xi(r)$ must change sign with $r$ at least once. 
They are systems with finely balanced positive and negative correlation. 
\end{enumerate}  
 
The distinction between \ref{1.} and \ref{2.} is typical of the   
statistical physics of critical phenomena in order to distinguish  
a critical state (case \ref{2.}) from a non-critical state (case \ref{1.}).  
In this context the concept of correlation length is central.  
The correlation length is a measure of the distance up to which
one has spatial memory of the spatial variations in the mass 
density \cite{huang}.
There is no unique definition of this length scale, but from a  
phenomenological point of view it can be defined as the length  
scale up to which the effect of a small local perturbation in   
the system is felt. This is due to the {\em fluctuation-dissipation theorem}  
which links the response of the system to a local perturbation and the   
large scale behavior of the two-point correlation function  
(for the different precise definitions of the correlation length see  
for example \cite{ma84}). A simple definition is (but see also  
\cite{gaite,gsld00}) 
\be 
r_{\rm{corr}}^2  
=\frac{\int_{\Omega} r^2 d^d r \; \;  
|\tilde \xi(r)|}{\int_{\Omega}  d^d r \; \; |\tilde \xi(r)|} \;. 
\label{corrlength} 
\ee 
In case \ref{1.} one can generally define a finite correlation length, 
while in case \ref{2.} it will generally diverge.  In particular  
in the case  $\xi(r)\sim \exp (-r/r_c)$, $r_c$ is indeed then the correlation 
length, while for a positive power-law $\xi(r) \sim 1/r^{\gamma}$ and 
$\gamma < d$ (case \ref{2.}) $r_{\rm{corr}} \rightarrow \infty$.

Case \ref{3.} is typical of ordered compact systems with small   
correlated  perturbations. One can meet this kind  
of correlation function for example   
in the statistical physics of liquids, glasses,  
phonons in lattices. The concept of correlation length in  
this context is less central, and the extension of its use to this class  
of systems is not particularly useful. Instead it is  
appropriate to classify the correlation properties of these 
systems directly through the integral of the correlation 
function as we have done. It is this behaviour of their correlations 
which distinguishes them from the other two cases, just as these 
cases are typically distinguished from one another by the 
value (finite or infinite) of their correlation length. 
Certainly, as we have noted, the use of the term ``correlation 
length'' in the cosmological literature, which is defined \cite{pee80} 
as a scale defining the amplitude of the correlation function, is 
in no way related to its use in statistical physics.  
 
Before continuing with a more detailed discussion of the nature of 
this class of {\em super-homogeneous} distributions to which standard  
cosmological models belong, we clarify one quite common  
misunderstanding about them in cosmology.

\section{$P(0)=0$ and constraints in a finite sample} 
 
As we have noted in the introduction the physical meaning of 
the constraint $P(0)$, equivalent to Eq. (\ref{super}), is often  
missed in the cosmological literature because of a confusion  
with the so-called ``integral constraint'', which is another 
very similar, but actually completed different, constraint.   
Let us clarify this point. 
 
The ``integral constraint'' refers in this context 
to a constraint which appears in the estimation of  
the correlation function in a finite sample ($S$, say).  
It is a constraint which indeed can take the superficially 
similar form to (\ref{super}):  
\be 
\int_S  d^3r  \; \; \tilde \xi_E(r)=0  
\label{int-constraint-e} 
\ee 
where the subscript indicates that the integral is over the  
finite sample volume, and $\tilde{\xi}_E(r)$ is the value  
of the {\it estimator} of the correlation function. This is  
in general a quantity calculable from the sample whose  
ensemble average converges to the real correlation function  
at any finite scale when the boundaries of the sample go  
to infinity.  
 
That such a constraint has in principle nothing  
to do with the constraint $P(0)=0$ is clear from the fact that 
it is one which holds independently of what kind of distribution 
the sample is taken from.  Its origin is  
simple, in the fact that the the mean mass density, relative 
to which fluctuations are estimated, is taken in the estimator 
from  the sample itself. Therefore, roughly speaking, the 
positive correlations measured relative to this density  
are constrained to be balanced by anti-correlations, giving rise to 
a constraint like Eq. (\ref{int-constraint-e}). More specifically  
the two point correlation function can be written as 
\be 
\tilde\xi(r)=\frac{\langle n(r) \rangle_p}{\langle n \rangle} -1  
\label{2ptcorr} 
\ee 
where $\langle n(r) \rangle_p$  
is the mean density at distance $r$ from an occupied 
point and $\langle n \rangle$  
the true (unconditional) mean density. Integrating  
this expression over the volume of the sample $S$ gives the relation 
\be 
\langle N_S \rangle_p - \langle N_S \rangle   
= \langle n \rangle  \int_S  d^3r  \; \;  \tilde\xi(r) 
\label{2ptcorr-int} 
\ee 
where $ \langle N_S \rangle_p$  
is the average number of points in the sample volume, 
with a point at the origin by construction, while $\langle N_S \rangle$  
is the  
average number of points is the same volume, 
but without the condition that there is a point at the location 
of the observer. If one estimates the true mean density 
(i.e. $\langle n \rangle = \langle N_S \rangle  
/ V_S$) in a galaxy catalogue sample  
from the actual density in the sample (i.e. on average 
$ \langle N_S \rangle_p/V_S$), the estimator for the correlation function  
will by construction on average obey the condition  
Eq.(\ref{int-constraint-e}) 
i.e. 
\be 
\int_S  d^3r  \; \; \langle \tilde\xi_E(r) \rangle =0.  
\label{int-constraint-aver} 
\ee 
Such a systematic (i.e. ensemble average) offset between the  
estimated and the real correlation function is sometimes 
referred to in the cosmological literature on the subject  
as ``bias''. It is only in very specific circumstances, with 
certain estimators, that Eq.(\ref{int-constraint-e}) holds for a 
single sample. For an estimator of the form \cite{slmp98} 
\be 
\tilde \xi_E(r)=\frac{\Gamma(r)}{n_S} -1  
\label{2ptcorr-est} 
\ee 
where $\Gamma(r)$ is estimator of $\langle n(r) \rangle_p$, and $n_S$ an  
estimate of the mean-density from the sample,  
Eq.(\ref{int-constraint-e}) will hold if   
\be 
n_S=\frac{1}{V_S} \int_S \Gamma(r) d^3r \;. 
\label{2ptcorr-est-cond} 
\ee 
This is in fact a perfectly good prescription for how to estimate 
the mean density in a finite sample, but one that is not used in 
most estimators, which typically have a variance around the 
average behaviour Eq. (\ref{int-constraint-aver}).  
Estimators \footnote{For a discussion of estimators used  
in the cosmological literature see e.g. \cite{kerscher}.}   
which do not take the mean density as the simple density of points 
in the sample do not in general obey even  
Eq.(\ref{int-constraint-aver}), but will always obey some  
constraint of this type, which cannot be avoided because it 
is intrinsic to the fact that any real sample contains an 
occupied point at the position of the observer.

In summary there are necessarily constraints on the correlation function  
$\tilde \xi_E(r)$ measured in a finite sample, which may take a form  
similar to the condition Eq. (\ref{super}) defining super-homogeneous   
distributions, but over a finite integration volume. These two 
kinds of constraint have a completely different origin and meaning, 
one describing an intrinsic property of the fluctuations in a certain 
class of distributions, the other a property of the estimated correlation 
function of any distribution as measured in a finite sample. Their formal  
resemblance however is not completely without meaning and can be understood 
as follows: in a super-homogeneous distribution the fluctuations between 
samples are extremely suppressed, being smaller than poissonian 
fluctuations; in a finite sample a similar behaviour is artificially 
imposed since one suppresses fluctuations at the scale of 
the sample by construction. An estimator which imposes the  
relation Eq.(\ref{int-constraint-aver}) on the estimated correlation  
function would therefore be expected to make a smaller error for  
the class of super-homogeneous distributions than for others.  
We will return to issues such as this in forthcoming work.

\section{Super-homogeneous Distributions} 
 
In this section we discuss the properties of long-range ordered  
mass distributions, or super-homogeneous distributions.  
We first discuss the simplest example of such a situation,  
represented by a lattice of particles.  It has many of the  
relevant properties under discussion. By studying  
its perturbation (the ``shuffled lattice'') it is possible to  
understand the properties of more isotropic distributions, 
both continuous and discrete, which are characterized again  
by long-range order. The main feature of these distributions,  
as we have discussed, is that $\sigma^2(R)$ (the unconditional variance)  
decays faster than in the uncorrelated Poissonian case, i.e. faster  
than $R^{-d}$(where $d$ is the space dimension). We then discuss how 
a continuous field with such correlations can be constructed, making 
it clear that the intuitions about the nature of the fluctuations 
in the shuffled lattice can be extended to the continuous case.  
We mention here that one physical model in which such correlations 
are found\footnote{We thank  
B. Jancovici for  describing  
these systems to us.}  
is the ``one component plasma'' studied, for example, in 
\cite{lebowitz,jankovici}. This models a Coulomb system of discrete positive 
charges in a continuous negatively charged background. In equilibrium 
the charges reach an extremely ordered glass-like configuration with 
PS at small $k$ like that of the shuffled lattice.

\subsection{The perfect lattice}                 
 
The microscopic density in the case of particles placed  
on the sites of a regular lattice (in any dimension)   
can be simply written as   
\be  
\label{lat1}  
\rho(\vec{r}) = \sum_{\vec{R}}   
\delta(\vec{r}-\vec{R}-\vec{\eta})  
\ee  
where $\vec{R}$ is the generic lattice displacement vector  
and $\vec{\eta}$ is position vector of the lattice site with $\vec{R}=0$  
with respect to the origin of coordinates,  
i.e. $\vec{R}+\vec{\eta}$  
runs over all the lattice sites.  For simplicity let us suppose  
we have a cubic lattice of unitary lattice spacing.   
Then, in order to eliminate the dependence on the position of  
the origin of coordinates with respect to the lattice,  
we can define an ensemble of lattices by varying the position of the   
origin with uniform probability in an unitary cell.  
For instance in $d=3$,    
$-1/2\le \eta_x,\eta_y,\eta_z< 1/2$ where $x,y,z$ are the axis coordinates,  
and then the ``ensemble average'' is $\left<...\right>\equiv  
\int\int\int_{-1/2}^{1/2}d\eta_xd\eta_yd\eta_z...$.    
  
Clearly we have that  
\be  
\label{lat2}  
\langle \rho(\vec{r})\rangle = 1 \;.  
\ee  
We want to compute now the two-point correlation function  
$\langle \rho(\vec{r}_1)\rho(\vec{r}_2)\rangle$.   
Firstly we have that   
\be  
\label{lat3}  
\rho(\vec{r}_1)\rho(\vec{r}_2) = \sum_{\vec{R}_1,\vec{R}_2}  
\delta(\vec{r}-\vec{R}_1-\vec{\eta})   
\delta(\vec{r}-\vec{R}_2-\vec{\eta}) \;.  
\ee  
from which one obtains  
\bea  
\label{lat5}  
\langle \rho(\vec{r}_1)\rho(\vec{r}_2) \rangle =  
%\\ \nonumber   
\sum_{\vec{R}} \delta  
\left( \vec{r}_1 - \vec{r}_2 - \vec{R} \right)  
\eea  
and hence the two-point correlation function   
\bea  
\label{lat6}  
\tilde \xi(\vec{r}_1,\vec{r}_2)=  
\frac{\langle \rho(\vec{r}_1)\rho(\vec{r}_2) \rangle } {\langle \rho \rangle^2}  
-1 =\\ \nonumber   
\! \left(\! \sum_{\vec{R}} \delta(\vec{r}_1 - \vec{r}_2 - \vec{R})  
\right)\!\! -\!1 \;.  
\eea  
Note that $\tilde\xi(\vec{r}_1,\vec{r}_2) \equiv   
\tilde\xi(\vec{r}_1 - \vec{r}_2)$ which means that our occupation stochastic  
process (i.e. the ensemble) is stationary.  
However since it is not invariant for generic spatial rotations we have   
$ \tilde\xi(\vec{r}_1 - \vec{r}_2)\ne \tilde\xi(|\vec{r}_1 -\vec{r}_2|) $  
(the lattice breaks spatial isotropy).   
  
In order to evaluate $P(\vec{k})$ we need to  
perform the FT  of $\tilde\xi(\vec{r})$.  
In the case of a lattice this gives simply  
\be  
\label{lat7}  
P(\vec{k})=   
\sum_{\vec{h} \ne 0} \delta(\vec{k} -\vec{h})  
\ee  
where the sum is extended to all the dual lattice vector  
$\vec{h}$ satisfying the {\em duality} condition  
$\vec{h} \cdot \vec{R} = 2 \pi m$, where $m$ is any integer,   
but with the exception of $\vec{h}=\vec{0}$.  
Note that, because of this last condition, also in this case $P(\vec{0}) = 0$.  
  
It has been shown \cite{kendall} that for the simple lattice 
in $d$ dimensions the fluctuations in a  
ball\footnote{ Note that \cite{kendall,beck}   
for the same quantity  
in cubic boxes of size $R$ one obtains  
$\lan M(R)^2 \ran - \lan M(R) \ran^2 \sim R^{d+1}$. This is a typical  
pathology of the lattice which is not a real stochastic particle distribution,  
having a {\em deterministic} discrete translation symmetry.   
This pathology is eliminated   
in the case we consider below of a ``shuffled lattice''.} of radius $R$,  
centered on a randomly chosen point,   
behaves as   
\be  
\label{lat10}  
\lan M(R)^2 \ran - \lan M(R) \ran^2 \sim R^{d-1}.  
\ee  
compared to the Poisson behaviour  
$\lan M(R)^2 \ran - \lan M(R) \ran^2 \sim R^d$.   
This result can be understood as follows:  
\begin{itemize}   
\item In the Poisson distribution, if we take two randomly placed spheres of  
same radius, the numbers of particle contained in them differ  
by an amount which is typically of the order of the square root of the   
average number ($\sim R^d$);  
\item In the case of a lattice the two numbers differ by an amount   
which corresponds to a Poissonian fluctuation (i.e. the square root)  
of the number of particles contained in the last shell of the sphere  
of thickness equal to the lattice spacing (which scales as $R^{d-1}$).  
\end{itemize}  
This is due to the strong order of the particles in the lattice from a   
large scale point of view. Thus the lattice has a behaviour of 
its mass variance which places it in the super-homogeneous category, 
with the limiting decreasing behaviour    
\be  
\label{lat11}  
\sigma^2(R) \sim R^{-(d+1)}\;,  
\ee  
i.e. $\sigma^2(R) \sim R^{-4}$ for $d=3$.

%%%%%%%%%%%%%%%%%%%%%%%%%%%%%%%% 

\subsection{The ``shuffled'' lattice}                 
\label{shuffled-lattice} 
  
In this section we define a super-homogeneous  
stochastic distribution of particles obtained  
from a lattice which shows more evident resemblances  
with the cosmological H-Z case.  
The recipe is the following:  
1) Consider a cubic lattice of particle, as discussed above;  
2) take a particle of the distribution and draw a lattice-oriented cubic box   
of size $l$ larger than the lattice spacing centered on the particle itself;  
3) displace the particle to a randomly chosen point of this box;  
4) repeat for each particle of the lattice.      
  
We can write $\rho(\vec{r})$ for a certain   
realization of this stochastic process as  
\be  
\label{slat1}  
\rho(\vec{r}) = \sum_{\vec{R}} \delta(\vec{r}-\vec{R}-  
\vec{\eta}_{\vec{R}}-\vec{\eta})\,,  
\ee  
where $\vec{R}$ and $\vec{\eta}$ have the same meaning as before,  
and $\vec{\eta}_{\vec{R}}$ is the vector  
giving the displacement of the particle  
in the box from the lattice rest position $\vec{R}+\vec{\eta}$.  
By definition each component of the vector $\vec{\eta}_{\vec{R}}$   
is a random number  
uniformly distributed in the interval $[-l/2,+l/2]$.  
Therefore, for instance in $d=3$, the ensemble average   
$\left<...\right>$ is now defined to be:  
\be  
\label{av-s-l}  
\left<...\right>=\int\int\int_{-1/2}^{1/2}d^3\eta\prod_{\vec{R}}  
\int\int\int_{-l/2}^{l/2}\frac{d^3\eta_{\vec{R}}}{l^3} ... \;. 
\ee  
  
After some algebra one finally obtains that   
$\tilde\xi(\vec{r}_1,\vec{r}_2)\equiv \tilde\xi(\vec{r}_1-\vec{r}_2)$  
(i.e. the ensemble is stationary) and in $d$ dimensions 
for integer $l$ one finds exactly:  
\be  
\label{slat4}  
\tilde\xi(\vec{r})  
= \delta(\vec{r}) -   
\prod_{k=1}^d \left\{\begin{array}{ll}  
\frac{1}{l} - \frac{|r_k|}{l^2} \; \; \mbox{if} \; \; |r_k| < l  
\\ \nonumber  
0  \; \; \mbox{if} \; \; |r_k| \ge  l  
\end{array}  
\right.  \;. 
\ee  
It is very simple to verify that $\int\int\int_{-\infty}^{\+\infty}  
\tilde\xi(\vec{r})d^3r =0$, which is the condition of super-homogeneity.  
Note that for $l\rightarrow +\infty$ $\tilde\xi(\vec{r})$ reduces   
correctly to the simple delta function, 
i.e. to the Poisson correlation function. In fact in that limit  
we must of course obtain a Poissonian distribution of particles  
without correlations.  
Note that consequently, 
we have the following non-commutativity of the limits:  
\be  
\label{limits}  
0=\lim_{l\rightarrow\infty}  
\int\int\int_{-\infty}^{\+\infty}\tilde\xi(\vec{r})  
\ne \int\int\int_{-\infty}^{\+\infty}\lim_{l\rightarrow\infty}\tilde  
\xi(\vec{r})=1\,.  
\ee  
  
One can also find an exact form of $P(\vec{k})$ by applying  
Eq.(\ref{lat7a}) in $d$-dimensions:  
\be  
\label{slat5}  
P(\vec{k}) = \frac{1}{(2\pi)^d} \left[ 1 - \prod_{i=1}^{d}   
\frac{2(1 -\cos(l\,k_i))} {l^2 \,  
k_i^2}  \right]\,.  
\ee   
Let us analyze the behavior for small values of $k$. At   
the leading order we can write  
\be  
\label{slat6}  
P(\vec{k}) \approx \frac{l^2}{(2\pi)^d} \frac{k^2}{12}\,,  
\ee  
which implies an isotropic behavior for $k\rightarrow 0$   
even though by construction   
$ P(\vec{k})$ is not isotropic for a general   
$\vec{k}$.   
Note that we have the $P(\vec{0})=0$ behaviour of the H-Z spectrum. 
Finally, the fact that for $k\rightarrow\infty$ the PS tends  
to a positive constant, means that such a distribution is Poissonian at small  
scale ($r<l$).  
  
It is now easy to calculate analytically the unconditional number variance.  
In particular the calculation can be done exactly in cubic boxes with 
the same   
lattice symmetry and, with some simple approximations, in spheres.  
In both cases one obtains $\sigma^2(R)\sim R^{-4}$ for large scales,  
as in the case of a lattice,  
but eliminating the pathology of different scaling behaviors between   
cubic boxes and spheres which we noted is present for the rigid lattice.   
Note  
that the fact that $\sigma^2(R)\sim R^{-4}$ at large scales  
corresponds to the fact that, despite the ``shuffling'' of  
particles with respect  
to the lattice, the strong lattice order is maintained at large scales. 
For non-integer $l>1$  , even though calculations are cumbersome
and $\xi(r)$ is not simply writtable, 
the main results about $P(k)$ for small $k$ and $\sigma^2(R)$ 
for large $R$ are the same.

The H-Z spectrum has this same behaviour characteristic of lattice-like 
order at large scales, while its small $k$ PS is $P(k) \sim k$  
instead of $\sim k^2$. This spectrum corresponds to more power 
at large scales. We will see in the next section that this can be  
associated with an appropriately more ordered (i.e. coherent) shuffling  
of the lattice, and precisely what kind of large scale correlations  
is required to obtain the H-Z spectrum will be made explicit. The crucial 
point is that such shuffling must leave intact at very 
large scales the strong order of the lattice, so that one still has 
the characteristic behaviors we have seen in the shuffled lattice  
(a correlation function which is negative at large scales and 
integrates to zero, a normalised variance in spheres decreasing faster than  
the volume). We thus say that the distribution described by the  
H-Z spectrum has a lattice-like or, more appropriately  
because of the isotropy, glass-like long range order. More 
specifically it can be characterized as a glass with superimposed  
opportune coherent long-range perturbative waves of displacement.

In relation to this description it is interesting to make some 
brief comments on cosmological N-body simulations  
(see e.g.\cite{jenkins98}).  
In this context a standard algorithm used to generate initial  
conditions for these simulations involves imposing perturbations  
on a perfect lattice (or sometimes even ``glassy'' configuration).  
At first sight this would suggest that the point we are making  
about the H-Z spectrum is in fact already understood in the  
cosmological literature, or at least in the part of it  
on N-body simulations. This is not  
the case, and it is worth explaining this to avoid any possible 
confusion on this point. 
This technique for generating initial conditions has  
in fact been introduced to avoid problems with 
Poisson noise at small scales (large $k$) in the discretization  
procedure, and not because the system being simulated is understood  
to actually intrinsically resemble a lattice at small $k$.  
Indeed the 
primary goal of most of these simulations has been to study  
the dynamical evolution in a range of scales well below  
the ``turn-over'' in the PS, where the PS (at large $k$)  
 has a negative power-law form ($\sim k^{-\beta}$, with 
$\beta > 0$, corresponding to a positive correlation function  
with a ``critical'' power law behaviour). Thus this procedure is  
applied primarily in a range of scales where the system being  
modeled does not intrinsically resemble a lattice or  
glass at all.  
 
Only the more recent very large simulations  
describe the larger scales at which the initial conditions  
should have $P(k) \sim k$. In this case too the use of a 
displaced lattice in setting up initial conditions is not 
because the underlying system is understood to be lattice-like, 
but is simply inherited as a numerical technique for the same 
small scale considerations \cite{white93}.  
Indeed, as will be further clarified 
in the following section, there is in principal no reason why one 
has to start from a lattice to produce such a spectrum; nor indeed 
is it certain that one obtains the right correlation properties 
if one starts from a lattice. What is true is that the spectrum 
of the initial conditions, if it is H-Z, should be glass-like in the 
sense we have discussed. A real space analysis of the initial  
conditions actually used in such simulations  
shows \cite{thierry} that they do not in fact have the  
appropriate properties. 
  
Note finally that glass-like systems belong to a wide 
family of distributions for which the common feature is 
that $P(k) \sim k^a$ with $a>0$ for $k \rightarrow 0$  
and hence $P(0)=0$. However such behaviors in the PS 
do not imply directly that $\xi(r)$ has a negative  
power-law tail at large scales. In particular this is 
not true if the PS has a singularity for $P(0) \ne 0$,  
as happens in many systems. For example \cite{thierry} 
the glass-like distributions (unperturbed and perturbed) 
used as initial conditions in cosmological N-body simulations 
have indeed an oscillating $\xi(r)$ at all scales, 
and a mass variance $\sigma^2(r) \sim r^{-4}$.  
Thus we emphasize that the negative power-law tail of 
the real-space correlation function of the H-Z distributions 
in cosmology is a very particular feature of 
these models.

\subsection{Uniform distributions with a displacement field}  
  
Let us consider the case of a mass distribution (a density  
field) obtained by superimposing a random displacement field  
on a completely uniform density field.  
  
Let the uniform density field be $\rho_0(\vec{r})=\rho_0$ and  
superimpose on it the stochastic displacement field $\vec{u}(\vec{r})$   
(the infinitesimal volume $dV$  
at $\vec{r}$ is displaced by $\vec{u}(\vec{r})$).  
Let us call $\rho(\vec{r})$ the resulting density field.  
We suppose that the stochastic field displacement is   
the realization of a stationary and isotropic  
stochastic process characterized by the  
probability density functional ${\cal P}[\vec{u}(\vec{r})]$.  
In this way ${\cal P}[\vec{u}(\vec{r})]$ defines also an ensemble   
of density fields $\rho(\vec{r})$ which is stationary and isotropic, 
with $\left<...\right>$ the ensemble average.  
  
By applying the mass conservation (i.e. the continuity equation) we find   
\be  
\frac{\rho(\vec{r})-\rho_0}{\rho_0}\simeq   
-\vec{\nabla}\cdot\vec{u}(\vec{r})\,.  
\label{dis}  
\ee  
If we call as usual $\xi(r)$ the reduced two-point correlation function  
of the density field we can write  
\be  
\xi(r)=\left<\vec{\nabla}\cdot\vec{u}(\vec{r})  
\vec{\nabla}\cdot\vec{u}(\vec{0})\right>\,.  
\label{xi-u}  
\ee  
Then, taking the FT of both sides of Eq.~(\ref{xi-u}),  
and making use of the statistical isotropy, we obtain  
\be  
P(k)\sim k^2 P_u(k)\,,  
\label{P-u}  
\ee  
where $P(k)$ is the usual PS of the mass density field  
and $P_u(k)$ is the PS of the displacement field.  
 
Since $P_u(k)$ is a itself the PS of a SSP it is 
subject to the constraints of the Khintchine theorem. Thus  
at small $k$ it must diverge slower than $\sim k^{-3}$,  
allowing to obtain at most $P(k) \sim k^{-1}$, corresponding  
to a real space correlation function which must 
go to zero faster than $1/r^2$ as $r\rightarrow\infty$.  
Therefore any continuous SSP of the ``substantially Poissonian''  
and ``super-homogeneous'' type can be obtained  
in this way,  but not all the ``critical'' type behaviors. 
In particular one can obtain a H-Z type spectrum with 
$P_u(k)\sim  k^{-1}$ describing a critical type SSP.  
 
What is the relation to the discrete case?  
If we suppose that $P_u(0)=const.>0$, i.e. the displacement field  
at large spatial scales is Poissonian (i.e. uncorrelated), we find 
that for $k\rightarrow 0$ one has $P(k)\sim k^2$. This is 
exactly the same asymptotic behaviour as that we found for the  
case of the shuffled lattice.  Indeed we obtained the latter 
through the superposition of an uncorrelated random  
displacement field to a ``uniform'' background, and thus the 
result is natural. Moreover in general we would expect the  
relation Eq.~(\ref{P-u}) to give us in the discrete case  
the large scale behaviour of a set of fluctuations imposed  
on a discretization of the continuous uniform density field,  
and in particular of a rigid lattice which is simply such an 
object. Thus if instead of shuffling the lattice as in the 
previous section we superimpose correlated fluctuations with 
a spectrum $\sim k^{-1}$ we will obtain at large scales 
a distribution with H-Z behaviour, and the associated real 
space properties at large scales. 
 
Let us return again finally to our comments at the end of the 
last section on N-body simulations (see \cite{thierry} 
for a more complete discussion of this point).  
With the previous construction 
it is easy to see why one arrives at the idea of generating 
a distribution with a certain PS by displacing 
points on a rigid lattice. The lattice is simply a discretization 
of the continuous uniform background which is then perturbed.  
One could in principle start from a Poissonian distribution  
of particles - for the generation of the large $k$ 
behavior - which can be always made ``uniform'' to any 
desired precision (assuming no practical limitation on the 
number of points used), and then superimpose the displacement 
field to produce the required PS. If the distribution  
produced is to be H-Z, and particluary of CDM type,  
it will have the ``super-homogeneity'' of a  
lattice at the corresponding scales, with the characteristic real space 
behaviors we have used to define it. Equally if one starts from a  
lattice one can arrive at distributions which are not super-homogeneous. 
Indeed, as we noted above, in the context of N-body simulations 
in cosmology the displacement from a lattice to produce initial 
conditions has been introduced in simulations describing the  
evolution in cosmological models at scales where the models  
are not H-Z, but rather have a positive correlation function  
with a ``critical'' power-law. The starting point of a lattice  
has been favored over a ``uniform'' Poisson distribution 
simply because of numerical limitations, the latter producing  
at any feasible resolution too much small scale noise overwhelming 
correlations at small scales.  In summary the central point we  
are making here is {\it not} that a H-Z type spectrum can be obtained  
in principal by perturbing a lattice; rather the crucial point is that  
such a system is intrinsically lattice-like, irrespective of  
how a discrete realization of it is constructed in practice. 
These are two completely different things.

%%%%%%%%%%%%%%%%%%%%%%%%%%%%%%%%%%%%%%%%%%%%%%%%%%%%%%%%%%%%%%% 
%%%%%%%%%%%%%%%%%%%%%%%%%%%%%%%%%%%%%%%%%%%%%%%%%%%%%%%%%%%%%%% 
%%%%%%%%%%%%%%%%%%%%%%%%%%%%%%%%%%%%%%%%%%%%%%%%%%%%%%%%%%%%%%% 

\section{Discussion and Conclusions}  
  
First we return to the use of several terms ``scale-invariance'' in  
cosmology. We have described in section(\ref{H-Zsection}) with what 
meaning this term has been introduced in cosmology: it refers to 
the fact that the variance of the mass (or equivalently gravitational 
potential) has an amplitude at the horizon scale which does not  
depend on time. The PS associated with this behaviour is 
that of a correlated system which is of the super-homogeneous type.  
This use of the term ``scale-invariance''  
therefore is not in any way analogous to its  
(original) use in statistical physics. In this context it is 
associated with a distinctly different class of distributions 
which have special properties with respect to scale  
transformations: typically critical systems, like a  
liquid-gas coexistence phase at the critical point,  
which have a well defined homogeneity scale and a  
reduced two-point  correlation function   
which decays as a non-integrable power law: $\xi(r) \sim r^{-\gamma}$  
with $0<\gamma<3$. In particular the term does not have  
anything to do with the {\it amplitudes} of fluctuations being  
independent of scale: the amplitudes of fluctuations vary with scale,  
while the system is correlated at all scales.  
 
We note that one might be tempted to associate the 
term ``scale-invariance''  
in the context of cosmology simply to the power-law 
behaviour of the correlation function. From a mathematical point of view,  
the terminology ``scale-invariance''  could be used for any  
distribution with a power-law tailed  
 $\xi(r)$, that is satisfying $\xi(r')=A(b) \xi(r)$  
for $r' = br$ . In this case the H-Z spectrum would, however, 
be no more ``scale-invariant'' than any other spectrum  
with a power-law form at small $k$.  
In terms of its meaning 
in physics however this usage is restricted  
to the context of critical phenomena, in which  
it has been demonstrated to be very powerful and useful.  
As we have discussed, this case  is completely different  
from the systems we have termed super-homogeneous. In particular  
in the former 
systems fluctuations are always large at all scales, 
which formally is associated with the non-integrability  
of the correlation function. The super-homogeneous systems 
have correlations of a completely different kind: 
they  are delicately balanced to make the mass fluctuations  
smaller than for a Poisson type distribution.

In section (IV) we have briefly discussed  the use of the term 
``correlation length''. Historically this term was introduced  
in the context of critical phenomena to characterize the  
transition towards a state of 
the system where (positive) correlations are long range and the 
normalized fluctuations of the mass (or of other extensive 
quantities) decay with the scale more slowly than in the non-critical 
Poissonian state. Moreover in this context,  
the meaning of correlation length is given through  
the fluctuation-dissipation theorem, in which the 
correlation length plays the role of  
the distance up to which the system responds to a 
local external perturbation. It is thus a scale used to 
capture the essential physical distinction between 
two types of distributions which are both in turn 
distinct from the superhomogeneous systems. The  
essential difference between each of these three cases 
can be best characterized according to the large scale  
behavior of the mass fluctuations in each case. 
It is therefore useful to take as classification parameter 
the value of the integral of $\tilde\xi (r)$ over  all space 
as we have done, while the notion of correlation length has 
no obvious or unique generalization in this case.

Another term whose meaning it is useful to clarify  
is ``fractal''.  Fractal distributions,  
which are the prototype of scale-invariant geometrical 
distributions,   
represent a more extreme case of   
correlated systems: their average density is zero in the   
infinite volume limit and the conditional  
average density $\left<n(\vec{r})\right>_p$  
decays to zero as a power law $r^{-\gamma}$ as a function 
of the distance from an occupied point (with $0<\gamma<d$).   
The fractal dimension of the mass distribution is given by  
$D_f=d-\gamma$ \cite{man82,falconer}.  
%There is a link between  
%this case and  the ``scale-invariant''  
%behaviors we have previously mentioned: 
%if we take a liquid-gas coexistence phase at the   
%critical point and define appropriately  
%a new density field through  the subset of   
%over-densities we obtain a fractal mass distribution.  
A fractal is  
inhomogeneous at all scales and the concept  
of average density in a finite sample centered on   
an occupied point has no intrinsic meaning,  
because it depends on the sample size.  
Moreover, since the asymptotic average density of a fractal  
distribution is zero, both $\xi(r)$ and $P(k)$ are undefined 
for such a system \cite{slmp98}. Instead one has to work directly  
with the unnormalized conditional density $\left<n(\vec{r})\right>_p$.  
In general,  before introducing the estimators of  
$\xi(r)$ and $P(k)$ for a finite sample of a system whose 
underlying properties one does not know (e.g. the distribution 
of galaxies in the Universe), one needs to verify  
that the estimator of the average density  $\rho_0$ is not  
strongly dependent on the sample size \cite{slmp98}.  
 
There is sometimes confusion in the cosmological literature 
about the meaning of ``fractal'' in connection with the notion of scale  
invariance. For example in the book by Peacock \cite{peacock}   
(Sect.16) the author writes that  
``The Zeldovich spectrum is a scale-invariant spectrum....''.  
It is then shown that the PS of the fluctuations  
in the gravitational potential $P_{\phi} \sim k^{-3}$ and hence   
the auto-correlation function, or the quantity  
$\Delta_{\phi}^2 \sim P_{\phi}(k) \cdot k^3$ is a constant:    
the author concludes that ``Since potential perturbations  
govern the flatness of space-time, this says that the scale-invariant   
spectrum corresponds to a metric that is {\bf fractal} \footnote{Bold   
font is author's.}: space-time has the same degree of `wrinkliness'  
of each resolution scale.'' Both the term ``fractal'' and  
``scale-invariance'' are used here in an incorrect and  
misleading way with respect to any of the meanings attached 
to them in their (original) context of statistical physics.  
In particular the H-Z spectrum does not have any properties  
which would allow it to be associated with a fractal 
mass distribution.

We have discussed the criterion which leads to the H-Z  
spectrum. We have pointed out that the usual naive 
formulation of this condition is in fact not satisfied 
by the spectrum $P(k) \sim k$, and that one must  
phrase the condition in terms of the variance in Gaussian 
spheres at the scale of the horizon. This is because 
the spectrum $P(k) \sim k$ is singled out by the  
constancy (as a function of time) of $k^3 P(k)$ at the  
horizon scale, which cannot be taken to be proportional  
to the variance in spheres of radius $R \sim k^{-1}$ 
for $P(k) \sim k^n$ and $n \geq 1$. We have emphasized 
that this is not at all an unphysical behaviour due 
to the ideality of a sphere. Rather it actually tells us  
something very fundamental about the nature of these 
distributions: They are distributions which are so 
ordered at large scales that the variance of mass 
at large scales really does come from small scales. 
The H-Z spectrum marks the transition to a 
pure lattice-like behaviour of the normalised 
variance in spheres $\sigma^2(R) \sim 1/R^4$, which 
has been shown to be the most rapid possible 
decay of this quantity for {\it any} stochastic distribution 
of points.  
 
What then does the use, widespread in cosmology, of a  
Gaussian sphere mean for such a distribution?  
%Is its use (widespread in cosmology) meaningful in this case? 
Mathematically it simply filters out the power up to a certain  
mode $k$ (which then dominates the integral). Physically it 
can be extremely misleading if interpreted incorrectly as 
a characterization of a variance in real space. Consider  
the example of the shuffled lattice. It has PS $\sim k^2$ at  
small $k$ (i.e. $kl \ll  1$, where $l$ is the shuffling scale). 
Using a Gaussian sphere one would infer that the variance  
at large scale goes as $1/R^5$. Physically we know that  
all the variance comes from small scales in this case, 
and that this behaviour dominated by the lower cut-off 
gives $\sigma^2(R) \sim 1/R^4$. The behaviour in Gaussian 
spheres comes from the fact that one is smearing 
the small scale behaviour over the scale $k^{-1}$. As the 
sphere grows the lower cut-off grows too. One would 
obtain the same behaviour by taking a different smearing 
scale on the sphere's boundary, but by making this scale 
change in proportion to the radius of the sphere. The  
behaviour observed has to do with the very particular 
way one is smearing, and the real space properties of 
the system are actually obscured. The only usefulness 
of the Gaussian sphere is an alternative way of  
saying that $P(k) \sim k^2$, i.e. as a statement about 
$k$ space properties, not real space ones. It does  
not describe in any useful way a property of  
the system in real space. In particular the H-Z 
criterion should be understood really as a $k$ space 
one, and caution should  be applied to its  
formulation as  ``constancy of mass variance at the  
horizon scale''.

We have highlighted the fact that all current cosmological models 
will share at large scales the characteristic behaviour in real 
space of the H-Z spectrum. Specifically we note primarily the 
very characteristic lattice-like behaviour of the variance  
in spheres  $\sigma^2(R) \sim R^{-4}$ (up to 
a small correction which is formally logarithmic for the case 
of exact H-Z), as well as the characteristic negative 
(non-oscillating) power-law tail in the two point correlation function 
$\xi(r) \sim -r^{-4}$. In this paper we have not addressed  
practical questions concerning the observation of such  
behaviour in cosmological data. In particular one would 
expect such behaviour to be seen in principle, if these models 
are correct, in the distribution of matter in the Universe 
at large scales, and in particular in the distribution of 
galaxies. So far such behaviour has not been observed.  
Rather the characteristic feature of galaxy clustering 
at small scales is that it shows fractal behaviour 
\cite{slmp98,rees99}, which as we have noted corresponds 
to a very different kind of distribution than that  
described by CDM type models. A centrally important (and 
much debated \cite{slmp98,rees99,joycesylos_ApJ2000}) 
observational question is the determination of the scale  
marking the transition from this behaviour to homogeneity. 
%Only once such a determination has been achieved does 
%it make sense to  
%As we have discussed, a fractal-like distribution 
%is very different from the critical system-like   
%power law correlation (around a well-defined  
%non-zero average density) described by  
%CDM models at small scales, at least in the linear regime.  
%In order to detect the correlations predicted by CDM  
%in the distribution of galaxies, one should  
%first find a clear crossover towords homogeneity 
%i.e. a scale beyond which the average density becomes  
%a well-defined (i.e. sample-independent)  
%concept \cite{joycesylos_ApJ2000,slmp98}. 
On much larger  scales galaxy structures 
should then present the super-homogeneous character  
of the H-Z type PS. Indeed this should be a critical  
test of the interpretation of measurements of CMBR  
in terms of the H-Z picture on large  
spatial scales \cite{cobe,boomerang,maxima}. 
%Clearly the link between the observed fractal properties 
%of the galaxy distribution and such super-homogeneous 
%temperature fluctuations is a central problem for 
%theoretical cosmology.  
Observationally a crucial 
question is the feasibility of measuring the transition 
between these regimes directly in galaxy distributions. 
With large forthcoming galaxy surveys it 
may be possible to do so, but this is a question which must 
address exactly the statistics of these surveys and the 
exact nature of the signal in any given model. These are 
questions we will address in future works. One other direct  
usage of the results developed here is in the context of  
N-body simulations, in which one studies numerically  
the evolution of perturbations in cosmological models, 
and a knowledge of their real space characteristics can 
be very useful. We refer to \cite{thierry} for a detailed  
discussion of this point.

\section*{Acknowledgments}    
We thank T. Antal, R. Ball, Y.V. Baryshev, T. Baertschiger,   
R. Durrer, P. Ferreira, B. Jancovici,  
L. Pietronero and F. Vernizzi for useful discussions and comments.  
This work has been partially supported by the     
EC TMR Network  ''Fractal structures and  self-organization''      
\mbox{ERBFMRXCT980183} and by the Swiss NSF.

%%%%%%%%%%%%%%%%%%%%%%%%%%%%%%%%%%%%%%%%%%%%%%%%%%%%%%%%%%%%%%%%  

\end{document}